\documentclass[
aps,pra,amsmath,amssymb,twocolumn,
floatfix,groupedaddress]{revtex4-2}
\usepackage{mathrsfs}
\usepackage{color}
\usepackage{graphicx}
\usepackage{dcolumn}
\usepackage{braket}
\usepackage{bm}
\usepackage{hyperref}
\usepackage{hypernat}
\usepackage{version}
\usepackage{epstopdf}
\usepackage{float}

\DeclareGraphicsExtensions{.eps,.ps,.png}

\DeclareMathOperator*{\SumInt}{%
\mathchoice%
  {\ooalign{$\displaystyle\sum$\cr\hidewidth$\displaystyle\int$\hidewidth\cr}}
  {\ooalign{\raisebox{.14\height}{\scalebox{.7}{$\textstyle\sum$}}\cr\hidewidth$\textstyle\int$\hidewidth\cr}}
  {\ooalign{\raisebox{.2\height}{\scalebox{.6}{$\scriptstyle\sum$}}\cr$\scriptstyle\int$\cr}}
  {\ooalign{\raisebox{.2\height}{\scalebox{.6}{$\scriptstyle\sum$}}\cr$\scriptstyle\int$\cr}}
}

\newcommand{\rmi}{\mathrm{i}}
\newcommand{\rme}{\mathrm{e}}

\begin{document}

\preprint{APS/123-QED}
\title{Time-dependent Hole States in Multiconfigurational Time-Dependent Hartree-Fock Approaches: Applications in Photoionization of Water Molecule}
\author{Zhao-Han Zhang$^{1}$}
\author{Yang Li$^{1}$}\email{liyang22@sjtu.edu.cn}
\author{Himadri Pathak$^{3,4}$}
\author{Takeshi Sato$^{5,6,7}$}
\author{Kenichi L. Ishikawa$^{5,6,7,8}$}
\author{Feng He$^{1,2}$}\email{fhe@sjtu.edu.cn}

\affiliation{
$^1$Key Laboratory for Laser Plasmas (Ministry of Education) and School of Physics and Astronomy,
Collaborative Innovation Center for IFSA (CICIFSA), Shanghai Jiao Tong University, Shanghai 200240, China\\
$^2$Tsung-Dao Lee Institute, Shanghai Jiao Tong University, Shanghai 201210, China\\
$^3$Quantum Mathematical Science Team, Division of Applied Mathematical Science, RIKEN Center for Interdisciplinary Theoretical and Mathematical Sciences (iTHEMS),  2-1 Hirosawa Wako, Saitama 351-0198, Japan\\
$^4$Computational Molecular Science Research Team, RIKEN Center for Computational Science (R-CCS), 7-1-26 Minatojima-minami-machi, Chuo-ku, Kobe, Hyogo 650-0047, Japan\\
$^5$Department of Nuclear Engineering and Management, Graduate School of Engineering, The University of Tokyo,7-3-1 Hongo, Bunkyo-ku, Tokyo 113-8656, Japan\\
$^6$Photon Science Center, Graduate School of Engineering, The University of Tokyo, 7-3-1 Hongo, Bunkyo-ku, Tokyo 113-8656, Japan\\
$^7$Research Institute for Photon Science and Laser Technology, The University of Tokyo, 7-3-1 Hongo, Bunkyo-ku, Tokyo 113-0033, Japan\\
$^8$Institute for Attosecond Laser Facility, The University of Tokyo, 7-3-1 Hongo, Bunkyo-ku, Tokyo 113-0033, Japan
}
\date{\today}

\begin{abstract}
By simulating the real-time multielectron wavefunction with the multi-configurational time-dependent Hartree-Fock (MCTDHF) approach, we conduct an \textit{ab initio} study of the single-photon ionization process of a body-fixed water molecule ($\mathrm{H_2O}$) driven by attosecond pulses. To this end, we present a full-dimensional implementation of the MCTDHF method based on one-center expansions, allowing for the simulation of arbitrarily polarized lasers and multi-center polyatomic potentials. With a rigorous definition of the time-dependent hole state (TDHS) using the time-domain generalization of extended Koopmans' theorem (TD-EKT), we derive the reduced ion density matrix within the MCTDHF framework, which inherently encodes the total and channel-resolved photoionization cross sections of $\mathrm{H_2O}$. The cross sections obtained are benchmarked against existing experimental and theoretical results, validating the TDHS formalism. Furthermore, by adjusting the phase delay and intensity ratio of a pair of orthogonally polarized attosecond pulses, we explore the ultrafast control of attosecond coherence between electronic states of $\mathrm{H_2O^+}$.
\end{abstract}


\maketitle
\section{Introduction}
The wavefunction-based multi-configurational time-dependent Hartree-Fock (MCTDHF) theory~\cite{lode2020colloquium,ishikawa2015review} has emerged as a general and effective framework for solving the multi-electron time-dependent Schr\"{o}dinger equation (ME-TDSE). As an \textit{ab initio} theory, it enriches the theoretical toolbox for the study of ultrafast electron dynamics in real-world systems, particularly in laser-driven many-electron atoms and molecules. The original formulation of the MCTDHF method represents the correlated time-dependent multi-electron wavefunction with a multi-orbital formalism, where both single-particle orbitals and configuration interaction coefficients evolve dynamically over time~\cite{zanghellini2003mctdhf,kato2004time,caillat2005correlated}. The approach achieves a balance between accuracy and computational efficiency by efficiently exploring a necessary subset of the multi-electron Hilbert space, thereby capturing the essential physics of a broad range of laser-driven processes.

Despite a wealth of successful applications in various contexts
\cite{caillat2005correlated,kato2008time,jordan2008core,sukiasyan2009multielectron,haxton2011multiconfiguration,
haxton2012single,hochstuhl2014time,sawada2016implementation}, 
the MCTDHF method adopts an all-electron complete active space (CAS) scheme, in which the number of electronic configurations grows exponentially with the number of electrons. For this reason, even though the method formally converges to the \textit{exact} solution of ME-TDSE, achieving full convergence is often computationally prohibitive. To address this challenge, generalized MCTDHF methods have been proposed, incorporating restricted active space (RAS) strategies. These advancements have led to the broader time-dependent multi-configurational self-consistent-field (TD-MCSCF) framework, which incorporates various approaches with a unified structure of equations of motion (EOMs). Notable examples include the time-dependent complete-active-space self-consistent-field (TD-CASSCF) method~\cite{sato2013time}, the time-dependent restricted-active-space self-consistent-field (TD-RASSCF) method~\cite{miyagi2013time,miyagi2014time}, and the time-dependent  occupation-restricted multiple-active-space (TD-ORMAS) method~\cite{sato2015time}. By elaborating the configuration spaces, these methods have significantly improved computational efficiency~\cite{sato2016time,omiste2018attosecond}, making the calculations practical for large systems with a limited number of active electrons. They have been successfully applied to a variety of problems, including few-photon ionization~\cite{li2016direct,omiste2017electron,omiste2019effects}, strong-field ionization, and high harmonic generation~\cite{tikhomirov2017high,orimo2019application,orimo2019comparison,li2024implementation}.

The rapid advancement of laser technology has enabled the generation of ultrashort extreme ultraviolet (XUV) pulses on the attosecond timescale~\cite{sansone2006isolated,chini2014generation,gaumnitz2017streaking,wang2024ultrashort}. When a neutral target is ionized by attosecond XUV pulses, the residual ion can occupy multiple symmetry-allowed states, leading to a multichannel photoionization process where quantum entanglement and coherence naturally arise in photofragments~\cite{vrakking2021control,koll2022experimental,ishikawa2023control,ruberti2024bell,laurell2025measuring}. Usually, the internal state of the photoion is difficult to measure directly as it often decays away before reaching the detector, whereas the emitted photoelectron is able to carry information about the ionization process to infinity, fulfilling the basis of photoelectron spectroscopy. However, for attosecond pulses, when the energy separation between different ionic states is small, it becomes challenging to determine which state the photoelectron is entangled with, making state-resolved analysis crucial for both experimental and theoretical investigations. As an important indicator to study the dynamics of ionization, photoelectron momentum distributions (PMD) are frequently calculated in TD-MCSCF simulations~\cite{haxton2012single,omiste2017electron,orimo2019application,orimo2019comparison,omiste2019effects}, where the total PMD is directly available from the expectation value of the electron density operator in momentum space. However, unlike theoretical frameworks that explicitly expand the wavefunction in terms of predefined ion states~\cite{clarke2018r}, or experimental setups that enable coincidence measurements~\cite{dorner2000cold}, there is no general framework within the TD-MCSCF method for extracting ion-state-resolved observables, such as ion-state-resolved PMDs. This limitation significantly restricts the applicability of the MC-TDSCF method to address the challenges of attosecond photoionization dynamics.

In the accompanying paper~\cite{Letter}, we systematically overcome this difficulty by generalizing the extended Koopmans' theorem to the time domain, where a definition of the time-dependent hole state (TDHS) naturally arises. As a general formalism, it is suitable for any TD-MCSCF methods, including TD-CASSCF, TD-RASSCF, TD-ORMAS, etc. By combining precise TDHS with projection approaches~\cite{madsen2007extracting,argenti2013photoionization} or the time-dependent surface flux (t-SURFF) method~\cite{Tao2012,scrinzi2012t}, we can obtain hole-state-resolved PMDs. In this work, we present a detailed description of the TDHS formalism. As a demonstration, we apply this method to the single ionization process of a body-fixed water molecule to validate this approach.

This paper is organized as follows. 
In Sec.~\ref{Theory}, we briefly review the formal theory of TD-MCSCF and derive the EOM of TDHS. Based on this, we systematically develop approaches to extract channel-resolved observables, with an emphasis on the channel-resolved momentum distribution and the reduced density matrix of photoion. 
In Sec.~\ref{Computational Approaches}, we present the spin-adapted numerical implementation of the non-relativistic MCTDHF theory, incorporating a gauge-invariant frozen-core extension. Our implementation employs a one-center expansion of the single-particle orbitals using radial B-spline functions and spherical harmonics, enabling the treatment of arbitrarily polarized laser pulses and polyatomic molecular potentials. To enhance computational efficiency, we introduce an angular discrete variable representation for the evaluation of the two-electron Coulomb term. 
In Sec.~\ref{Results}, we apply our theory and numerical implementation to extract both the total and partial photoionization cross sections of a body-fixed water molecule. Good agreement is achieved with existing theoretical and experimental values, validating the accuracy of our approach. Furthermore, by analyzing the reduced ion density matrix, the control of the quantum coherence of the water molecular ion with a pair of orthogonally polarized attosecond pulses is also investigated.\\
\section{Theory of Time-dependent Hole States}\label{Theory}
\subsection{EOM of TD-MCSCF}
The TD-MCSCF approach is devoted to the numerical solution of the ME-TDSE
\begin{equation}\label{ME-TDSE}
i|\dot{\Psi}(t)\rangle = \hat{H}(t)|\Psi(t)\rangle,
\end{equation}
where the time-dependent Hamiltonian has a general second-quantized form
\begin{equation}\label{Hamiltonian}
\begin{aligned}
\hat{H}(t) &= \sum_{\mu\nu} h_{\mu\nu}(t)\hat{a}^\dag_\mu(t)\hat{a}_\nu(t) \\
&+ \frac{1}{2}\sum_{\mu\nu\lambda\tau} g_{\mu\nu\lambda\tau}(t)\hat{a}^\dag_\mu(t)\hat{a}^\dag_\nu(t)\hat{a}_\lambda(t)\hat{a}_\tau(t).
\end{aligned}
\end{equation}
The time-dependent operators $\hat{a}^\dagger_\mu(t)$ and $\hat{a}_\mu(t)$ create and annihilate an electron in the time-dependent single-electron state $\phi_\mu(t)$, respectively. We use Greek letters $(\mu,\nu,\lambda,\tau,\cdots)$ for the indices of single-electron states. The one-body integral $h_{\mu\nu}(t)$ incorporates the kinetic terms, the laser-electron interaction, and the nuclei-electron Coulomb interaction. The two-body integral $g_{\mu\nu\lambda\tau}(t)$ accounts for the electron-electron interactions.

In the TD-MCSCF method, the ME-TDSE is solved approximately using the multi-configurational wavefunction \textit{ansatz}
\begin{subequations}\label{ansatz-general}
\begin{align}
|\Psi(t)\rangle&=
\sum_{I} C_I(t)|I(t)\rangle, \\
|I(t)\rangle &= 
\prod_{\mu}
(\hat{a}^\dagger_{\mu}(t))^{I_{\mu}}|0\rangle.
\end{align}
\end{subequations}
Here, the integer array $I$ specifies the occupation numbers of the single-electron states in the time-dependent electron configuration $|I(t)\rangle$. The size of $I$ is equal to the number of electrons $N$ and $I_\mu\in\{0,1\}$. A practical calculation only incorporates a finite number of $\phi_\mu$, $\mu\in\mathcal{O}$. The single-electron Hilbert space is thus separated into two orthogonal parts, the occupied and unoccupied spaces, spanned by $\{\phi_\mu\}_{\mu\in\mathcal{O}}$ and $\{\phi_\mu\}_{\mu\notin\mathcal{O}}$, respectively. Those $\mu\notin\mathcal{O}$ will be excluded from the summation in Eq.~\eqref{ansatz-general}.

Under the orthonormal condition $\langle\phi_\mu|\phi_\nu\rangle=\delta_{\mu\nu}$, the EOMs for the expansion coefficients $C_I$ and orbitals $\phi_\mu$ are derived from the time-dependent variational principle (TDVP)~\cite{dirac1930note,frenkel1934wave,mclachlan1964variational}, yielding the following coupled nonlinear differential equations (for simplicity, we omit the explicit time dependence in the notation in the following)
\begin{subequations}\label{EOM}
\begin{align}
\label{EOM:CI}
i\dot{C}_I &= \sum_{J}\langle I|\hat{H}-\hat{R}|J\rangle C_J, \\
\label{EOM:OB}
i|\dot{\phi}_\mu\rangle &= 
\hat{R}|\phi_\mu\rangle + \hat{Q}\hat{F}|\phi_\mu\rangle.
\end{align}
\end{subequations}
Here, $\hat{Q}$ is the projection operator onto the unoccupied space. $\hat{R}$ is an auxiliary operator to remove the redundant degree of freedom in the \textit{ansatz}, whose matrix elements, $R_{\mu\nu}\equiv\langle\phi_\mu|\hat{R}|\phi_\nu\rangle \equiv i\langle\phi_\mu|\dot{\phi}_\nu\rangle$ with $\mu,\nu\in\mathcal{O}$, need to be designated in advance. The redundancy arises from the arbitrariness of distributing time-dependent global phases between orbitals and CI coefficients without altering physics~\cite{meyer1990multi,caillat2005correlated}. In the standard all-electron MCTDHF theory~\cite{zanghellini2003mctdhf,kato2004time,caillat2005correlated}, $R_{\mu\nu}$ can be an arbitrary time-dependent Hermitian matrix, while for the various RAS extensions of MCTDHF, such as TD-CASSCF~\cite{sato2013time}, TD-RASSCF~\cite{miyagi2013time} and TD-ORMAS~\cite{sato2015time}, $R_{\mu\nu}$ should be determined by special procedures. The generalized Fock operator $\hat{F}$ is defined as
\begin{equation}\label{Def:Fop}%
\hat{F}|\phi_\mu\rangle\equiv\bigg(\hat{h}|\phi_\mu\rangle+\sum_{\nu\lambda\tau\in\mathcal{O}}\overline{\rho}_{\mu\nu\lambda\tau}\hat{g}_{\nu\lambda}|\phi_\tau\rangle\bigg),
\end{equation}
with matrix elements $F_{\mu\nu}=\langle\phi_\mu|\hat{F}|\phi_\nu\rangle$~\cite{sato2013time}.
Here, $\hat{h}$ is the one-electron Hamiltonian, and $\hat{g}_{\nu\lambda}$ characterizes the Coulomb potential excited by the electrons. Formally, we have
\begin{subequations}
\begin{align}
\hat{h}		 &=
\sum_{\mu\nu}h_{\mu\nu}\hat{a}^\dag_{\mu}\hat{a}_{\mu},\\
\hat{g}_{\nu\lambda}&=
\sum_{\mu\tau} g_{\mu\nu
\lambda\tau}\hat{a}^\dag_{\mu}\hat{a}_{\tau},
\end{align}
\end{subequations}
and
\begin{subequations}
\begin{align}
\overline{\rho}_{\lambda\mu\nu\tau} &= \sum_{\xi\in\mathcal{O}}\rho^{-1}_{\lambda\xi}\rho_{\chi\mu\nu\tau}, \\
\rho_{\mu\nu} 		&= \langle\Psi|\hat{a}^\dag_{\mu}\hat{a}_{\nu}|\Psi\rangle, \\
\rho_{\lambda\mu\nu\tau} 	&= \langle\Psi|\hat{a}^\dag_{\mu}\hat{a}^\dag_{\nu}\hat{a}_{\lambda}\hat{a}_{\tau}|\Psi\rangle.
\end{align}
\end{subequations}
The last two equations correspond to the one-body reduced density matrix (1RDM) and the two-body reduced density matrix (2RDM), respectively.

By solving Eq.~\eqref{EOM}, one obtains a multielectron wavefunction as an approximate solution of the ME-TDSE. The convergence of the method is controlled by the size of the occupied space. At the full configuration interaction (FCI) limit, the \textit{ansatz} is \textit{complete}, and $\Psi$ converges to the \textit{exact} solution of ME-TDSE. When the number of orbitals equals the number of electrons, the \textit{ansatz} reduces to the time-dependent Hartree-Fock (TDHF) method.
\subsection{EOM of TDHS}
For the consistency with TD-MCSCF theory, we construct the $(N-1)-$electron state by a linear combination of the single-vacancy states $|\Psi_\mu\rangle\equiv\hat{a}_\mu|\Psi\rangle$, $\mu\in\mathcal{O}$:
\begin{equation}
\label{Def:TDHS}
|\gamma\rangle = \sum_{\mu\in\mathcal{O}}|\Psi_\mu\rangle Z_{\mu\gamma},
\end{equation}
where $Z_{\mu\gamma}$ is a matrix of undetermined coefficients. The state $|\gamma\rangle$ is a one-hole state, which we call the TDHS. We assume that $|\gamma\rangle$ satisfies the projected ME-TDSE
\begin{equation}\label{PROJ-ME-TDSE}
i\hat{\varPi}|\dot{\gamma}\rangle=\hat{\varPi}\hat{H}|\gamma\rangle,
\end{equation}
where $\hat{\varPi}$ is the projection operator of the $(N-1)-$electron Hilbert space spanned by all single-vacancy states. This yields the following set of coupled differential equations for each $\mu\in\mathcal{O}$
\begin{subequations}\label{PROJ-ME-TDSE2}
\begin{align}
\label{PROJ-ME-TDSE-LHS}
&\sum_{\nu\in\mathcal{O}}i\big(\langle\Psi|\hat{a}^\dag_\mu\hat{a}_\nu|\dot{\Psi}\rangle+\langle\Psi|\hat{a}^\dag_\mu\dot{\hat{a}}_\nu|\Psi\rangle\big)Z_{\nu\gamma} + i\rho_{\mu\nu}\dot{Z}_{\nu\gamma}\\
\label{PROJ-ME-TDSE-RHS}
&=\sum_{\nu\in\mathcal{O}}\big(\langle\Psi|\hat{a}^\dag_\mu\hat{a}_\nu\hat{H}|\Psi\rangle+\langle\Psi|\hat{a}^\dag_\mu[\hat{H},\hat{a}_{\nu}]|\Psi\rangle\big)Z_{\nu\gamma}.
\end{align}
\end{subequations}
Using Eqs.~\eqref{Helper1} and \eqref{Helper2} (see Appendix~\ref{Relations}), Eq.~\eqref{PROJ-ME-TDSE2} is eventually simplified to
\begin{equation}
\label{EOM:TDHS}
-\sum_{\nu\tau\in\mathcal{O}}(F_{\nu\tau}-R_{\nu\tau})\rho_{\mu\tau}Z_{\nu\gamma} = \sum_{\nu\in\mathcal{O}} i\rho_{\mu\nu}\dot{Z}_{\nu\gamma},
\end{equation}
which is the EOM of TDHS. By introducing the $\rho$-weighted matrix
\begin{equation}
\tilde{F}_{\nu\mu}=\sum_{\tau\in\mathcal{O}}(F_{\nu\tau}-R_{\nu\tau})\rho_{\mu\tau},
\end{equation}
the EOM of TDHS has a simpler form of $i\rho \dot{Z}=-\tilde{F}^TZ$.

It is beneficial to show that the stationary version of Eq.~\eqref{EOM:TDHS} reproduces the extended Koopmans' theorem. 
If we impose the stationary condition $i\dot{Z}=Z\mathcal{E}_\gamma$, $i|\dot{\gamma}\rangle=|\gamma\rangle E_\gamma$, and $i|\dot{\Psi}\rangle=|\Psi\rangle E_0$ to Eq.~\eqref{PROJ-ME-TDSE2}, it follows that $\rho Z (E_{\gamma}-E_0-\mathcal{E})=0$. Therefore, $\mathcal{E}_\gamma=E_{\gamma}-E_0$ is the energy difference between $|\Psi\rangle$ and $|\gamma\rangle$, known as the ionization energy. By imposing the same stationary condition to Eq.~\eqref{EOM:TDHS}, it follows the extended Koopmans' theorem (EKT)\cite{koopmans1934zuordnung,day1974generalization,smith1975extension,morrell1975calculation}:
\begin{equation}\label{EKT}
-\tilde{F}^T Z=\rho Z\mathcal{E}.
\end{equation}
This explains why the ionization energy in TD-MCSCF should be described by EKT, as observed in Ref.~\cite{hochstuhl2014time}.
We note that $\tilde{F}$ is Hermitian if and only if $\dot{\rho}=0$, as evident from the EOM of 1RDM:
\begin{equation}\label{EOM:RDM1}
\begin{aligned}
i\dot{\rho}_{\mu\nu}
 &= \langle\Psi|[\hat{a}_\mu^\dag a_{\nu},\hat{H}]|\Psi\rangle + \langle\Psi|i\dot{\hat{a}}_\mu^\dag\hat{a}_\nu+i\hat{a}_\mu^\dag\dot{\hat{a}}_\nu|\Psi\rangle\\
 &= (\tilde{F}^T - \tilde{F}^*)_{\mu\nu}.
\end{aligned}
\end{equation}
We then prove the orthonormality of TDHS.
As $Z(0)$ is set as the $\rho$-orthonormal eigenvectors of Eq.~\eqref{EKT}, $Z^\dag \rho Z = I$ is automatically satisfied at $t=0$. Due to Eq.~\eqref{EOM:RDM1}, we have
\begin{equation}\label{EOM:orth}
i\frac{d}{dt}(Z^\dag \rho Z) = Z^\dag(-\tilde{F}^T + \tilde{F}^*+i\dot{\rho})Z=0, 
\end{equation}
indicating that the orthonormality is held at any time $t$.
\subsection{Total Photoelectron Momentum Distribution}
Following photoionization, the multi-electron system is torn apart and fragments will move to infinity. The PMD carries abundant information about the dynamics near the interaction region.
Various theoretical approaches have been raised for extracting it, among which the most rigorous approach is directly projecting the final state onto the exact multi-electron scattering state immediately after the end of the laser~\cite{madsen2007extracting,argenti2013photoionization}.
However, acquisition of the exact scattering states is often more complicated than the propagation itself.
As a practical alternative, it is possible to compute PMD by projecting the departing wavepackets onto plane waves or outgoing Coulomb waves after a sufficiently long time ($t_f$) until the photoelectron has reached the asymptotic region $(r>r_0)$. 

We will denote the combined spin and momentum coordinate by $\bm{\kappa}=(\bm{k}\sigma)$, where $\bm{k}$ is the asymptotic kinetic momentum of the electron and $\sigma$ is its spin polarization species. The operator $\hat{a}^\dag_{\bm{\kappa}}$ creates an electron in the asymptotic region with momentum $\bm{k}$ and spin polarization $\sigma$. In practice, the state $\hat{a}^\dag_{\bm{\kappa}}|0\rangle=|\bm{\kappa}\rangle$ is often realized by a masked plane wave state $|\bm{\kappa}\rangle=\hat{\Theta}|\chi_{\bm{\kappa}}\rangle$, where 
\begin{equation}\label{Def:Theta}
    \hat{\Theta}=\sum_{\mu\nu}\Theta_{\mu\nu}\hat{a}^\dag_\mu\hat{a}_\nu,\quad\Theta_{\mu\nu}=\langle\phi_\mu|\hat{\Theta}|\phi_\nu\rangle
\end{equation}
is the density operator in the exterior region (usually defined by a unit step function for single-electron states) and $\chi_{\bm{\kappa}}$ is a plane wave with spin polarization $\sigma$. The total PMD is defined as the differential probability density
\begin{equation}\label{Def:PMD}
\begin{aligned}
\mathcal{P}(\bm{\kappa}) 
&=\langle\Psi|\hat{a}^\dag_{\bm{\kappa}}\hat{a}_{\bm{\kappa}}|\Psi\rangle\\
&=\sum_{\mu\nu\in\mathcal{O}}\rho_{\mu\nu}\phi^*_\mu(\bm{\kappa})\phi_\nu(\bm{\kappa}),
\end{aligned}
\end{equation}
where $\phi_\mu(\bm{\kappa})=\langle\bm{\kappa}|\phi_\mu\rangle$. We mention that while Eq.~\eqref{Def:PMD} theoretically includes both single and multiple ionization contributions, in practice, and particularly in this work, single ionization dominates, making it a good approximation for the single-ionization PMD.
\subsection{Channel-Resolved Projection Approaches}\label{CRPMD1}
The introduction of TDHS allows us to define channel-resolved PMD, where the state of the photoelectron and the photoion are resolved simultaneously.
We will discuss two choices of the ion state $|\gamma\rangle$ for the projection approach, i.e., the TDHS and the field-free HS.
For a given reference ion state $|\gamma\rangle$, the projection $\mathcal{Q}_\gamma(\bm{\kappa})=\langle\gamma|\hat{a}_{\bm{\kappa}}|\Psi\rangle$ is always a linear combination of $\phi_\mu(\bm{\kappa})$, namely
\begin{equation}\label{Def:Q}
\mathcal{Q}_\gamma(\bm{\kappa}) = \sum_{\mu\in\mathcal{O}} D_{\gamma\mu}\phi_\mu(\bm{\kappa}).
\end{equation}
Hence, the channel-resolved PMD associated with state $|\gamma\rangle$ is defined as $\mathcal{P}_\gamma(\bm{\kappa})=|\mathcal{Q}_\gamma(\bm{\kappa})|^2=|\langle\gamma|\hat{a}_{\bm{\kappa}}|\Psi\rangle|^2$. If $\gamma$ is chosen as the TDHS defined by Eq.~\eqref{Def:TDHS} and evolved according to Eq.~\eqref{EOM:TDHS}, the coefficient matrix $D$ is actually the inverse of the matrix $Z$:
\begin{equation}
\label{Relation:D&Z}
D_{\gamma\mu}=\langle\gamma|\hat{a}_\mu|\Psi\rangle=\sum_{\nu\in\mathcal{O}}(Z^\dag)_{\gamma\nu}\rho_{\nu\mu}=(Z^{-1})_{\gamma\mu}.
\end{equation}
Hence, its EOM can also be derived,
\begin{equation}\label{EOM:D}
i\dot{D}_{\gamma\mu} = \sum_{\nu\in\mathcal{O}}(F_{\mu\nu}-R_{\mu\nu})D_{\gamma\nu},
\end{equation}
which is useful for extracting channel-resolved PMD. 

Alternatively, $\gamma$ can be chosen as the field-free HS given by Eq.~\eqref{EKT}. The corresponding projection is denoted by $\mathcal{Q}^\mathrm{F}_{\gamma}(\bm{\kappa})$, and the coefficient matrix $D$ should be replaced by $D^\mathrm{F}$, where
\begin{equation}\label{Field-Free-D}
D^\mathrm{F}_{\gamma\mu}(t) 
= \sum_{\nu\in\mathcal{O}}(Z^\dag(0))_{\gamma\nu}\langle\Psi(0)|\hat{a}^\dag_\nu(0) \hat{a}_\mu(t)|\Psi(t)\rangle.\\
\end{equation}
Combining Eqs.~\eqref{Def:TDHS}\eqref{Relation:D&Z}\eqref{Field-Free-D}, the self-correlation functions of TDHS can be expressed by $D^\mathrm{F}$ and $D$ alone,
\begin{equation}
    \mathcal{C}_{\gamma\gamma'}(t)=\langle\gamma(0)|\gamma'(t)\rangle = \sum_{\mu\in\mathcal{O}}D^\mathrm{F}_{\gamma\mu}(t)D^{-1}_{\gamma'\mu}(t),  
\end{equation}
suggesting a fundamental relation between the two kinds of channel-resolved PMD
\begin{equation}
    \mathcal{Q}^\mathrm{F}_{\gamma}(\bm{\kappa})=\sum_{\gamma'}\mathcal{C}_{\gamma\gamma'}\mathcal{Q}_{\gamma'}(\bm{\kappa}).
\end{equation}
We also mention that the \textit{sum condition}
\begin{equation}\label{SumCond}
\mathcal{P}(\bm{\kappa})=\sum_\gamma\mathcal{P}_\gamma(\bm{\kappa})
\end{equation}
requires that $D^\dagger D=\rho$. For TDHS, we have $D=Z^\dagger\rho$, so that the requirement is automatically satisfied. Other choices, including field-free HS, will generally fail this requirement. 
\subsection{Channel-Resolved t-SURFF Approaches}\label{CRPMD2}
One significant drawback of projection approaches is the requirement of a large simulation region to retain the electron wavepackets before the projection is performed. Although the complexity of the present MCTDHF implementation only scales linearly with the simulation radius, an efficient evaluation of PMD remains highly desirable. A good candidate is the t-SURFF approach~\cite{Tao2012,scrinzi2012t}, where the projection is turned into a time-integrated surface flux evaluation, thus eliminating the necessity to store the entire outgoing wavepacket. 

As an advantage of TDHS, the channel-resolved projection onto TDHS can be effectively integrated with t-SURFF, yielding a channel-resolved PMD method that we refer to as channel-resolved t-SURFF. The central idea of this method is to derive an EOM for $\mathcal{Q}_{\gamma}(\bm{\kappa})$. According to Eqs.~\eqref{EOM:OB}\eqref{Def:Q}\eqref{EOM:D}, we have
\begin{equation}\label{EOM:Q}
i\dot{\mathcal{Q}}_{\gamma}(\bm{\kappa}) = \sum_{\mu\in\mathcal{O}} D_{\gamma\mu}J_\mu(\bm{\kappa}),
\end{equation}
where the right-hand side is the channel-resolved surface flux given by
\begin{equation}
J_\mu(\bm{k}) \equiv \langle\chi_{\bm{\kappa}}|\hat{\Theta}\hat{F}-\hat{H}_V\hat{\Theta}|\phi_\mu\rangle\approx\langle\chi_{\bm{\kappa}}|[\hat{\Theta},\hat{H}_V]|\phi_\mu\rangle,
\label{SurfaceFlux}
\end{equation}
where $\hat{H}_V$ is the Volkov Hamiltonian, and $\chi_{\bm{\kappa}}$ is a Volkov solution in the presence of external laser field which ends up to a plane wave with momentum $\bm{k}$ and spin $\sigma$. Implementing the continuum as a screened Volkov solution ensures the gauge invariance of PMD, as $\hat{\Theta}$ maintains its form under local gauge transform. When the Coulomb terms play a negligible role, for example, the t-SURFF boundary $(r=r_0)$ is extremely large, the channel-resolved t-SURFF approach produces identical results to the projection approach. For practical use, the Coulomb terms are approximately neglected in t-SURFF, retaining only the commutator $[\hat{\Theta},\hat{H}_V]$, which is zero except at the t-SURFF boundary. The validity of this approximation in TD-MCSCF is demonstrated in Ref.~\cite{orimo2019application}.

Given that the surface flux will vanish at infinity time, the \textit{unitarity} of $\mathcal{P}_\gamma(\bm{\kappa})$ is strictly guaranteed. Combining this property with Eq.~\eqref{SumCond}, the unitarity of the entire PMD is also proved.
\subsection{EOM of time-dependent Dyson orbitals}
As an analog to Eq.~\eqref{Def:Q}, we define the projection in real space
\begin{equation}\label{Def:Qr}
\mathcal{D}_\gamma(\bm{\varrho}) = \langle\gamma,\bm{\varrho}|\Psi\rangle =\sum_{\mu\in\mathcal{O}} D_{\gamma\mu}\phi_\mu(\bm{\varrho}),
\end{equation}
where $\bm{\varrho}$ is the collection of spatial coordinates and spin.
$\mathcal{D}_\gamma(\bm{\varrho})$ and $\mathcal{Q}_\gamma(\bm{\bm{\kappa}})$ are interpreted as the time-dependent Dyson orbitals in coordinate and momentum representations, respectively, which is a conceptual generalization of the static Dyson orbital.
With the EOM of $D$, the EOM of the underlying state,
\begin{equation}
|\psi_\gamma\rangle=\sum_{\mu\in\mathcal{O}} D_{\gamma\mu}|\phi_\mu\rangle,
\end{equation}
can be derived
\begin{equation}
i|\dot{\psi}_\gamma\rangle = \hat{F}|\psi_\gamma\rangle.
\end{equation}
Note that $\psi_\gamma$ does not have unit normalization except in the Hartree-Fock limit.

\subsection{Photoion and Photoelectron Reduced Density Matrix}\label{RDM}
In multi-channel processes, the fragments are typically in mixed states. Therefore, a quantum theory of the photoion or photoelectron naturally involves the density matrix formalism. Recently, there has been an emergent interest in the quantum coherence and quantum entanglement property of the photoelectron and photoion~\cite{vrakking2021control,koll2022experimental,ishikawa2023control,ruberti2024bell,laurell2025measuring}, which is quantitatively characterized by the photoion or photoelectron's reduced density matrix (iRDM or eRDM). With a well-defined TDHS, we are now able to extract iRDM and eRDM from the TD-MCSCF theory. 

We start from the total density operator for the ionized part,
\begin{equation}
\hat{\rho}=\mathcal{N}^{-1}\SumInt_{\gamma\gamma',\bm{\kappa}\bm{\kappa}'} \mathcal{Q}_\gamma(\bm{\kappa})|\gamma,\bm{\kappa}\rangle\langle\gamma',\bm{\kappa}'|\mathcal{Q}^*_{\gamma'}(\bm{\kappa}'),
\end{equation}
where $\mathcal{N}$ is a normalization factor to be given later. By taking the partial trace over the electronic or ionic degree of freedom, we get the density operator of the photoelectron and the photoion,
\begin{subequations}
\begin{align}
    \hat{\rho}^{(\rme)}&=  \SumInt_{\bm{\kappa}\bm{\kappa'}} \rho^{(\rme)}(\bm{\kappa},\bm{\kappa}')|\bm{\kappa}\rangle\langle\bm{\kappa}'|,  \\
    \hat{\rho}^{(\rmi)}&=  \SumInt_{\gamma\gamma'} \rho^{(\rmi)}(\gamma,\gamma')|\gamma\rangle\langle\gamma'|,
\end{align}
\end{subequations}
where the matrix elements are
\begin{subequations}\label{iRDM&eRDM}
\begin{align}
\label{eRDM}
\rho^{(\rme)}(\bm{\kappa},\bm{\kappa}') &= \mathcal{N}^{-1}\SumInt_\gamma\mathcal{Q}_\gamma(\bm{\kappa})\mathcal{Q}^*_\gamma(\bm{\kappa}'),\\
\rho^{(\rmi)}(\gamma,\gamma') &= \mathcal{N}^{-1}\SumInt_{\bm{\kappa}} \mathcal{Q}_{\gamma}(\bm{\kappa})\mathcal{Q}^*_{\gamma'}(\bm{\kappa}).
\label{iRDM}
\end{align}
\end{subequations}
In the above definition, the dynamic phases are attributed to $|\gamma\rangle\langle\gamma'|$ and $|\bm{\kappa}\rangle\langle\bm{\kappa}'|$, hence $\rho^{(\rmi)}(\gamma,\gamma')$ and $\rho^{(\rme)}(\bm{\kappa},\bm{\kappa}')$ will approach stationary at infinite time.
By enforcing $\mathrm{Tr}\rho^{(\rmi)}=\mathrm{Tr}\rho^{(\rme)}=1$, the normalization factor is determined as
\begin{equation}
\mathcal{N}=\SumInt_{\bm{\kappa}} \mathcal{P}(\bm{\kappa})=\sum_{\mu\nu\in\mathcal{O}}\rho_{\mu\nu}\Theta_{\mu\nu},
\end{equation}
which is actually the average photoelectron number in the exterior region. Eqs.~\eqref{eRDM} and \eqref{iRDM} can be simplified to 
\begin{subequations}\label{iRDM&eRDM2}
\begin{align}
\label{eRDM2}
\rho^{(\rme)}(\bm{\kappa},\bm{\kappa}') &= \mathcal{N}^{-1}\sum_{\mu\nu\in\mathcal{O}}\rho_{\mu\nu}\phi^*_\mu(\bm{\kappa}')\phi_\nu(\bm{\kappa}),\\
\rho^{(\rmi)}(\gamma,\gamma') &= \mathcal{N}^{-1}\sum_{\mu\nu\in\mathcal{O}}D^*_{\gamma'\mu}D_{\gamma\nu}\Theta_{\mu\nu}.
\label{iRDM2}
\end{align}
\end{subequations}
We note that the calculation of eRDM does not need any knowledge of TDHS, while such knowledge becomes essential for iRDM. Due to the orthonormality of spin-orbitals, the matrix elements $\Theta_{\mu\nu}=\delta_{\mu\nu}-\Theta^<_{\mu\nu}$ can be uniquely determined by the information in the interior region. Therefore, iRDM does not require retaining the wavepacket in the exterior region, which significantly reduces the numerical cost. 

\section{Computational Approaches}\label{Computational Approaches}
The approach described in this section is a gauge-invariant generalization of the standard all-electron MCTDHF, and is closely related to the aforementioned RAS extensions of MCTDHF.

We employ the spin-adapted formalism in practical computations, which is advantageous for closed-shell systems where spatial orbitals are always identical for the two spin species. We introduce the notation $\sigma\in\{\alpha,\beta\}$ where $\alpha,\beta$ are two possible spin species, and assign the Latin letters $(i,j,k,\cdots)$ for the indices of spatial orbitals. The Greek indices will thus be replaced according to $\mu=(i_\mu,\sigma_\mu)$, and further simplified to $(i\sigma)$. The common spatial part shared by $\phi_{i\alpha}$ and $\phi_{i\beta}$ is denoted by $\phi_{i}$.
The computational approach classifies the occupied orbitals as active or frozen, denoted by $\{\phi_i\}_{i\in\mathcal{A}}$ and $\{\phi_i\}_{i\in\mathcal{C}}$, respectively. The size of the active (frozen) space is defined as $N_{\rm a}=|\mathcal{A}|$ $(N_{\rm c}=|\mathcal{C}|)$, and there are totally $\mathcal{N}_{\rm o}=N_{\rm c}+N_{\rm a}$ occupied orbitals. The number of $\alpha$- and $\beta$-electrons is denoted by $N_{\alpha}$ and $N_{\beta}$, respectively.\\
\subsection{Lasers}
We describe the laser field using classical electromagnetic theory, modeling the laser pulses as coherent light. The dipole approximation is adopted throughout this work, as the wavelength of the laser field significantly exceeds the spatial extent of the interaction region by more than two orders of magnitude. Both the velocity gauge (VG) and the length gauge (LG) are implemented.
The vector potential $\bm{A}$ in VG is transformed into
\begin{subequations}
\label{laser}
\begin{align}
A_{\pm 1} &= \mp\frac{1}{\sqrt{2}}(A_x\pm iA_y),\\
A_0 &= A_z,
\end{align}
\end{subequations}
where the components satisfy $A_\mu=(-1)^\mu A^*_{-\mu}$ for $\mu=-1,0,+1$. 
The unit vector $\bm{e}_{\mu}$ is defined in the same manner so that $\bm{e}_{\mu}^*\cdot\bm{e}_{\nu}=\delta_{\mu\nu}$ and $\bm{A}=A_x \bm{e}_x + A_y \bm{e}_y + A_z\bm{e}_z= A^*_{-1}\bm{e}_{-1}+A_0^*\bm{e}_0+A^*_{+1}\bm{e}_{+1}$.
$A_{-1}$ and $A_{+1}$ correspond to the right-handed and left-handed components, respectively. In LG, the electric field components are denoted as $E_\mu$, derived from the vector potential via $\bm{E}=-d\bm{A}/dt$ to ensure gauge invariance.\\
\subsection{Potentials}
To describe the polyatomic potential, we use the general one-center expansion for the Coulomb interaction
\begin{equation}\label{OCEpoten}
V(\bm{r}) = V_0(r) + \sum_{\lambda=0}^{+\infty}\sum_{\mu=-\lambda}^{\lambda} \sqrt{\frac{4\pi}{2\lambda+1}}Y_\lambda^\mu(\theta,\varphi) V^{\mu\lambda}(r).
\end{equation}
Here, the central part $V_0(r)=-Z_0/r$ represents the Coulomb potential arising from the nucleus located at $r=0$, while the contributions from all off-center nuclei are included in the multipolar terms $V^{\mu\lambda}(r)$, explicitly given by
\begin{equation}
\label{multipolarV}
V^{\mu\lambda}(r) = \sum_{d=1}^{N_{\rm pole}} -Z_d\sqrt{\frac{4\pi}{2\lambda+1}}Y_\lambda^{\mu*}(\Theta_d,\Phi_d)\frac{r_<^\lambda}{r_>^{\lambda+1}}.
\end{equation}
Here, $Z_d$ and $(R_d,\Theta_d,\Phi_d)$ are the charge and spherical coordinates of the $d$-th off-center atomic nucleus, respectively. $r_<=\min(r,R_d)$ and $r_>=\max(r,R_d)$ denote the smaller and larger radii between the electron and the nucleus, respectively.\\
\subsection{Multiconfiguration Wavefunctions}
In this work, frozen-core orbitals are always doubly occupied. All remaining orbitals are treated as active, forming a complete active space (CAS) in which any electron occupation is permitted. This classification significantly reduces computational complexity while preserving essential physical insights, enabling efficient simulations of ultrafast electron dynamics in molecular systems. 

Our spin-adapted \textit{ansatz} corresponding to Eq.~\eqref{ansatz-general} is
\begin{subequations}\label{ansatz}
\begin{align}
|\Psi\rangle&=
\sum_{I\in\text{CAS}} C_I|I\rangle, \\
|I\rangle &= 
\prod_{i\in \mathcal{C}}\hat{a}^\dag_{i\alpha}\hat{a}^\dag_{i\beta}
\prod_{j\in \mathcal{A}}
(\hat{a}^\dag_{j\alpha})^{I^\alpha_{j}}
(\hat{a}^\dag_{j\beta})^{I^\beta_{j}}|0\rangle.
\end{align}
\end{subequations}
Here, the spin-adapted formalism naturally separates the occupations into spin-specific substrings $I^{\sigma}$ for the occupation of $\sigma$-electrons. The spin-adapted EOM for the CI coefficients has the same form as Eq.~\eqref{EOM:CI}.
\subsection{Spin-adapted Operators}
As the spin is conserved due to $[\hat{S}_{z},\hat{H}]=0$, the spin indices in the Hamiltonian Eq.~\eqref{Hamiltonian} are traced off to give
\begin{equation}
\begin{aligned}
\hat{H}&= \sum_{ij} h_{ij}\sum_{\sigma}\hat{a}^\dag_{i\sigma}\hat{a}_{j\sigma} \\
 &+\frac{1}{2}\sum_{ijkl} g_{ijkl}\sum_{\sigma\sigma'}\hat{a}^\dag_{i\sigma}\hat{a}^\dag_{j\sigma'}\hat{a}_{k\sigma'}\hat{a}_{l\sigma'}.
\end{aligned}
\end{equation}
The spin-free one- and two-electron integrals are
\begin{subequations}
\begin{equation}
h_{ij}(t) = \int d^3\bm{r} \phi^*_i(\bm{r},t)\mathcal{H}\phi_j(\bm{r},t),
\end{equation}
\begin{equation}
\begin{aligned}
g_{ijkl}(t) = \int d^3\bm{r}d^3\bm{r}' \frac{\phi^*_i(\bm{r},t)\phi^*_j(\bm{r}',t)\phi_k(\bm{r}',t)\phi_l(\bm{r},t)}{|\bm{r}-\bm{r}'|},
\end{aligned}
\end{equation}
\end{subequations}
where the one-electron Hamiltonian $\mathcal{H}$ is
\begin{subequations}
\begin{align}
\mathcal{H}^{\rm LG}&=-\frac{1}{2}\bm{\nabla}^2 + \bm{E}(t)\cdot\bm{r} + V(\bm{r}), \\ 
\mathcal{H}^{\rm VG}&=-\frac{1}{2}\bm{\nabla}^2 - i\bm{A}(t)\cdot\bm{\nabla} + V(\bm{r})
\end{align}
\end{subequations}
in LG and VG, respectively. Note that the gauge invariance of MCTDHF theory is ensured by construction.

For tightly bound core electrons, the gauge-invariant frozen-core approximation is adopted~\cite{sato2016time}, so that we do not need to solve their orbitals from an EOM, but directly set their value by:
\begin{subequations}\label{FC}
\begin{align}
\phi^{\rm LG}_i(\bm{r},t) &=                                      \phi_i(\bm{r},0)\text{  for   }i\in\mathcal{C},\\
\phi^{\rm VG}_i(\bm{r},t) &= \exp\bigg(i\bm{A}(t)\cdot\bm{r}\bigg)\phi_i(\bm{r},0)\text{  for   }i\in\mathcal{C}.
\end{align}
\end{subequations}
The initial values of the frozen orbitals $\left\{\phi_i(\bm{r},0)\right\}_{i\in\mathcal{C}}$ are set to the canonical Hartree-Fock orbitals obtained in field-free Hartree-Fock calculations using the same method.

For active orbitals, the spin-adapted EOM has almost the same form as Eqs.~\eqref{EOM}, except that the reduced density matrices should be replaced by the spin-traced version, i.e.,
\begin{subequations}
\begin{align}
\overline{\rho}_{ijkl} &= \sum_{i'\in\mathcal{O}}\rho^{-1}_{ii'}\rho_{i'jkl}, \\
\rho_{ij} 		&= \sum_{\sigma}\langle\Psi|\hat{a}^\dag_{i\sigma}\hat{a}_{j\sigma}|\Psi\rangle, \\
\rho_{ijkl} 	&= \sum_{\sigma\sigma'}\langle\Psi|\hat{a}^\dag_{i\sigma}\hat{a}^\dag_{j\sigma'}\hat{a}_{k\sigma'}\hat{a}_{l\sigma}|\Psi\rangle.
\end{align}
\end{subequations}
We have merged the core-active interaction into an effective one-body operator $\hat{f}$, so that the generalized Fock operator is equivalently reduced to
\begin{equation}\label{Def:Fop}%
\hat{F}|\phi_i\rangle\equiv\bigg(\hat{f}|\phi_i\rangle+\sum_{jkl\in\mathcal{A}}\overline{\rho}_{ijkl}\hat{g}_{jk}|\phi_l\rangle\bigg).
\end{equation}
The effective one-body operator,
\begin{equation}
\hat{f}=\sum_{ij}f_{ij}\sum_{\sigma}\hat{a}^\dag_{i\sigma}\hat{a}_{j\sigma},
\end{equation}
now includes the usual one-body terms, the Coulomb and exchange interactions from the frozen-core, namely,
\begin{equation}
f_{ij} = h_{ij} + \sum_{k\in \mathcal{C}}\bigg(2g_{ikkj}-g_{ikjk}\bigg).
\end{equation}
In this work, we set the matrix elements $R_{ij}$ within the active orbital space to zero, whereas those connecting active and frozen orbitals are determined from their instantaneous values as~\cite{sato2016time}: 
\begin{equation}
\label{def:R}
R_{ij} = i\langle\phi_i|\dot{\phi}_j\rangle=R_{ji}^*,\quad i\in\mathcal{A},j\in\mathcal{C}.
\end{equation}
\subsection{Discretization}
To discretize the EOMs, the time-dependent orbitals are expanded via the one-center expansion
\begin{equation}
\label{monocentricEXP}
\phi_i(\bm{r},t) = \sum_{mln} c_{imln}(t) \frac{B_{n}(r)}{r}Y_{l}^{m}(\theta,\varphi).
\end{equation}
Here, the radial basis $B_n(r)$ is a set of B-spline functions~\cite{bachau2001applications} of the same order on a given knot sequence, and $Y_l^m(\theta,\varphi)$ is the spherical harmonics. In our notation, we denote by $(n|\mathcal{P}|n')$ the integral by sandwiching an operator $\mathcal{P}$ between $B_n(r)$ and $B_{n'}(r)$. Bold uppercase(lowercase) letters are used for matrices(vectors) in the spherical-B-spline basis. For brevity, we introduce a combined index $\iota$ to abbreviate the triplet $(m,l,n)$ when referring to their matrix elements.

To address the stiffness in the resulting discretized orbital EOMs, the right-hand side is separated into parts:
\begin{equation}
i\dot{\bm{c}}_i = \bm{S}^{-1}\bm{T}\bm{c}_i -\sum_{j\in\mathcal{O}} \bm{c}_j X_{ji} + \bm{Q}\bm{u}_i.
\end{equation}
Here, $\bm{S}_{\iota\iota'}=\delta_{mm'}\delta_{ll'}(n|n')$ is the overlap matrix of the basis, and $\bm{T}_{\iota\iota'}=\delta_{mm'}\delta_{ll'}(n|-\nabla^2/2+V_0|n')$ is the stiff part of the one-particle Hamiltonian matrix.
In this work, interaction terms arising from the laser and the off-center atoms are considered to be non-stiff.
The orbital-mixing matrix is defined as $X_{ji}=T_{ji}-R_{ji}$ with $T_{ji}=\bm{c}^\dag_j\bm{T}\bm{c}_i$.
The orbital-projection matrix, expressed formally as $\bm{Q}=\bm{S}^{-1}-\sum_{j\in\mathcal{O}}\bm{c}_j\bm{c}_j^\dag$, does not need to be stored in practical calculations.
The vector $\bm{u}_i$ collects all remaining terms, including the laser interaction, the off-center atomic potential, the core-active interaction, and the active-active interaction:
\begin{equation}\label{defu}
\begin{aligned}
\bm{u}_i &= \big(\bm{V} + \bm{W}\big)\bm{c}_i + \sum_{j\in \mathcal{C}} 2\bm{G}_{jj}\bm{c}_i-\bm{G}_{ji}\bm{c}_j \\
&+ \sum_{jkl\in\mathcal{A}} \overline{\rho}_{ijkl}\bm{G}_{jk}\bm{c}_l,
\end{aligned}
\end{equation}
where
\begin{subequations}
\label{def:MatrixElements}
\begin{align}
\bm{V}_{\iota\iota'} &= \sum_{\lambda=0}^{+\infty} \sum_{\mu=-\lambda}^{\lambda} C^{\mu\lambda}_{mm',ll'} V^{\mu\lambda}_{nn'},\\
\bm{W}^{\rm VG}_{\iota\iota'} &= -\sum_{\mu=-1}^{1} A^*_\mu\big(C^{\mu 1}_{mm',ll'}D_{nn'}+\overline{C}^{\mu 1}_{mm',ll'}L_{nn'}\big),\\
\bm{W}^{\rm LG}_{\iota\iota'} &= \sum_{\mu=-1}^{1} E^*_\mu C^{\mu 1}_{mm',ll'}R_{nn'},\\
(\bm{G}_{jk})_{\iota\iota'} &=  \sum_{\lambda=0}^{+\infty} \sum_{\mu=-\lambda}^{\lambda} C^{\mu\lambda}_{mm',ll'}G^{\mu\lambda}_{jk,nn'}.
\label{defG}
\end{align}
\end{subequations}
The radial integrals in Eq.~\eqref{def:MatrixElements} are given by 
\begin{subequations}
\begin{align}
D_{nn'}&=(n|i\frac{d}{dr}|n'),\\
L_{nn'}&=(n|\frac{1}{r}|n'),\\
R_{nn'}&=(n|r|n'),\\
V_{nn'}^{\mu\lambda}&=(n|V^{\mu\lambda}(r)|n'),\\
G^{\mu\lambda}_{jk,nn'}&=(n|G^{\mu\lambda}_{jk}(r)|n').
\end{align}
\end{subequations}
Here, $V^{\mu\lambda}(r)$ is given in Eq.~\eqref{multipolarV}. $G^{\mu\lambda}_{jk}(r)$ is the radial component (see Appendix~\ref{TwoElectronTerm}) accompanying $Y_\lambda^\mu(\theta,\varphi)$ in the one-center multipole expansion of the inter-electron Coulomb interaction arising from the source term $\phi_j^*(\bm{r})\phi_k(\bm{r})$.
Using 3-j symbols, we write the angular integrals in Eq.~\eqref{def:MatrixElements} in a compact form
\begin{equation}
\begin{aligned}
&C^{\mu\lambda}_{mm',ll'} = (-1)^m \sqrt{(2l+1)(2l'+1)}\\
&\times\bigg(
\begin{matrix}
l & \lambda & l'\\
0 & 0       & 0 \\
\end{matrix}
\bigg)\bigg(
\begin{matrix}
l & \lambda & l'\\
-m& \mu     & m'\\
\end{matrix}
\bigg),
\end{aligned}
\end{equation}
which satisfies $C^{\mu\lambda}_{mm',ll'}=(-1)^\mu C^{-\mu\lambda}_{m'm,l'l}$.
The symbol $\overline{C}^{\mu 1}_{mm',ll'}=i^{l'-l}\max(l,l')C^{\mu 1}_{mm',ll'}$ is also introduced.

As the EOMs of MCTDHF inherently conserve the symmetry properties of the orbitals (a detailed proof can be found in the appendix of Ref.~\cite{omiste2018attosecond}), the expansion scheme in Eq.~\eqref{monocentricEXP} should be the same for the orbitals inside the same symmetry group but could be different between symmetry groups in principle. For consistency, the same design in our previous single-electron work~\cite{zhang2023qpc} is adopted for all orbitals. 
We define $N_{\rm b}$ as the number of independent radial B-spline functions, $N_{\rm l}$ as the number of different angular quantum numbers $l$ for each magnetic quantum number $m$, and $N_{\rm m}$ as the total number of different $m$ for each orbital. As the orbital functions of deeply bound electrons oscillate rapidly near the central atom, we specifically fine-tune the B-spline knot distribution in this region to accurately capture this character.
If the laser polarization aligns with the mirror plane of the molecule, it saves significant numerical efforts to adopt the $\rm C_s$ point group. Specifically, we may set the mirror plane as the $x$-$y$ plane, so that the terms in the expansion Eq.~\eqref{monocentricEXP} vanish depending on whether $l-m$ is odd or even, as dictated by symmetry. In more general situations without evident symmetry, we implement the full expansion.

Although the one-center expansion in Eq.~\eqref{monocentricEXP} is notably efficient for atomic systems, its convergence for molecular orbitals tends to be considerably slower. Consequently, large values $N_{\rm l}$ and $N_{\rm m}$ are often indispensable for accurately describing bound molecular orbitals~\cite{toffoli2002convergence}.
Nevertheless, if all off-center atoms are hydrogen atoms and relatively close to the expansion center, reliable results can still be achieved with moderate choices of $N_{\rm l}$ and $N_{\rm m}$, typically around 20. This enables practical real-time simulation for small molecules such as $\rm NH_3$, $\rm H_2O$, $\rm CH_4$ near equilibrium geometry. When the total number of basis ($N_{\rm m}N_{\rm l}N_{\rm b}$) grows up to $10^4$, severe bottlenecks could arise in computing the two-electron Coulomb integral at each time step. To address this, we introduce an efficient computational strategy in Appendix.~\ref{TwoElectronTerm}.

In cases with large numbers of active orbitals, typically a few tens, two additional computational bottlenecks emerge, i.e., the evaluation of reduced density matrices and the matrix-vector product in the EOMs of CI coefficients. In this work, we limit the number of active orbitals to no more than 10, and there are no more than a few thousand electronic configurations in the CI expansion. We will discuss the treatment of larger CI space in future work.\\

\subsection{Propagation}
The discretized EOMs are propagated by Krogstad's ETD-RK4 solver~\cite{krogstad2005generalized,kidd2017exponential}. Here, the time-independent term $\bm{S}^{-1}\bm{T}$ is regarded as the only stiff part. We have verified that merging the monopole component of $\bm{G}_{jj}$ into $\bm{T}$ improves numerical stability, particularly when the orbital $\phi_j$ contains a deeply bound state localized around the central atom.  
A more rigid treatment of the stiffness problem involving the $\bm{A}\cdot\bm{\nabla}$ term can be found in Ref.~\cite{li2021implementation}. 
The $\varphi$-functions in the ETD-RK4 solver are implemented with the Lanczos-Arnoldi algorithm~\cite{saad2003iterative}. Throughout this work, the dimension of the Krylov subspace is fixed to $30$ for all partial waves during propagation. 

For the EOMs of $C_I$, no stiffness will occur, thus the algorithm degenerates to the conventional RK4 solver. 
The right-hand side of Eq.~\eqref{ansatz} is given by
\begin{equation}\label{CImatrix}
\langle I|\hat{H}-\hat{R}|J\rangle = \langle I|\hat{H}_a|J\rangle,
\end{equation}
where
\begin{equation}
\begin{aligned}
\hat{H}_a&=\sum_{ij\in\mathcal{A}}f_{ij}\sum_\sigma \hat{a}^\dag_{i\sigma}\hat{a}_{j\sigma} \\
&+ \frac{1}{2}\sum_{ijkl\in\mathcal{A}} g_{ijkl}\sum_{\sigma\sigma'}\hat{a}^\dag_{i\sigma}\hat{a}^\dag_{j\sigma'}\hat{a}_{k\sigma'}\hat{a}_{l\sigma}.
\end{aligned}
\end{equation}
It is suggested~\cite{sato2013time} that the diagonal entries of the CI matrix in Eq.~\eqref{CImatrix} should be offset by $-\langle\Psi|\hat{H}_a|\Psi\rangle$ to further suppress the numerical instability during propagation.

The propagation method can operate in either real or imaginary time. Real-time propagation (RTP) addresses problems driven by laser fields, while imaginary time propagation (ITP) is employed to generate the initial states in a self-consistent manner.\\
\subsection{Channel-Resolved Momentum Distribution}\label{PMD}
In spin-adapted formalism, the invariance under spin inversion guarantees $\mathcal{P}(\bm{k}\alpha)=\mathcal{P}(\bm{k}\beta)$. For simplicity, one often uses the spin-adapted PMD, $\mathcal{P}(\bm{k})\equiv\mathcal{P}(\bm{k}\alpha)+\mathcal{P}(\bm{k}\beta)=2\mathcal{P}(\bm{k}\alpha)$, which is explicitly evaluated by
\begin{subequations}
\begin{align}
\label{Cal:PMD}
\mathcal{P}(\bm{k}) &= 
\sum_{ij\in\mathcal{A}}\rho_{ij}(t_f)\phi^*_i(\bm{k},t_f)\phi_j(\bm{k},t_f),\\
\label{Def:MomOrb}
\phi_i(\bm{k},t_f) &= 
\frac{1}{(2\pi)^{\frac{3}{2}}} \int d^3\bm{r} e^{-i\bm{k}\cdot\bm{r}}\Theta(r-r_0)\phi_i(\bm{r},t_f).
\end{align}
\end{subequations}
The numerical method for evaluating Eq.~\eqref{Def:MomOrb} is quite similar to single-electron problems, and further details are omitted here.

The single-vacancy Hilbert space decouples into two degenerate subspaces distinguished by spin orientation. We denote the pair of degenerate hole states in these subspaces by
\begin{subequations}
\label{Def:IonSpinSubSpace}
\begin{align}
|\gamma\alpha\rangle &\in \text{span}\{\hat{a}_{i\alpha}|\Psi\rangle|i\in\mathcal{A}\},\\
|\gamma\beta \rangle &\in \text{span}\{\hat{a}_{i\beta}|\Psi\rangle|i\in\mathcal{A}\}.
\end{align}
\end{subequations}
Given spin-inversion symmetry, we isolate and redefine the spin-irrelevant part in the projection as 
\begin{subequations}
\begin{align}
\langle\gamma\alpha|\hat{a}_{\bm{k}\alpha}|\Psi\rangle&=\langle\gamma\beta |\hat{a}_{\bm{k}\beta}|\Psi\rangle\equiv\mathcal{Q}_\gamma(\bm{k}),\\ 
\langle\gamma\beta |\hat{a}_{\bm{k}\alpha}|\Psi\rangle&=\langle\gamma\alpha|\hat{a}_{\bm{k}\beta}|\Psi\rangle=0,
\end{align}
\end{subequations}
such that $\mathcal{P}_\gamma(\bm{k})\equiv2|\mathcal{Q}_\gamma(\bm{k})|^2=\mathcal{P}_{\gamma\alpha}(\bm{k}\alpha)+\mathcal{P}_{\gamma\beta}(\bm{k}\beta)$. Similarly, we redefine the identical parts in $D$ as
\begin{subequations}
\begin{align}
D_{\gamma\alpha, i\alpha}&=D_{\gamma\beta, i\beta}\equiv D_{\gamma i},\\
D_{\gamma\alpha, i\beta} &=D_{\gamma\beta, i\alpha}=0.
\end{align}
\end{subequations}
With these spin-adapted notations, the final expression of the spin-adapted channel-resolved PMD is 
\begin{equation}
\mathcal{P}_\gamma(\bm{k})= 2\sum_{ij\in\mathcal{A}} \phi^*_{i}(\bm{k},t_f) D^*_{\gamma i}(t_f)D_{\gamma j}(t_f)\phi_j(\bm{k},t_f).
\end{equation}
For both TDHS and field-free HS, the expression for $D_{\gamma i}$ is almost identical to the spin-unadapted theory in Eqs.~\eqref{EOM:D} and ~\eqref{Field-Free-D}, except that the Greek subscripts are replaced by Latin ones. In this work, Eq.~\eqref{EOM:D} is integrated by an explicit exponential integrator simultaneously with the integration of MCTDHF EOMs. At each time step, we read $F-R$ from MCTDHF and evaluate
\begin{equation}
D_\gamma(t+\tau) \approx \exp\{-i\tau(F(t)-R(t))\} D_\gamma(t).
\label{ExponentialIntegrator}
\end{equation}
The initial condition for Eq.~\eqref{ExponentialIntegrator} is given by $D(0)=Z(0)^{-1}$, where $Z(0)$ is solved from the spin-adapted EKT at $t=0$,
\begin{equation}
-\sum_{j\in\mathcal{A}} (F_{ji}-R_{ji})Z_{j\gamma} = \frac{1}{2}\sum_{j\in\mathcal{A}}\rho_{ij}Z_{j\gamma}\mathcal{E}_\gamma.
\label{spin-adapted-EKT}
\end{equation}

Similarly to the projection approach, the spin-adapted channel-resolved t-SURFF approach has an almost identical formalism except that the Greek letters are replaced by Latin ones. The numerical implementation of the surface flux term Eq.~\eqref{SurfaceFlux} is quite similar to the single-electron problem, so we will not repeat any more. Usually, the t-SURFF approach is combined with absorption boundary conditions. In this work, we will explicitly indicate instances whenever an absorber is utilized to eliminate outgoing wavepackets.\\  
\subsection{Channel-Resolved Single Ionization Yield}
Generally, the ionization yield for a neutral system refers to the fraction of specific atomic or ionic species present after ionization, e.g. $\rm X$, $\rm X^{+}$, $\rm X^{2+}$, etc. In Ref.~\cite{sato2013time}, the yield of $\rm X^{n+}$ is equivalently defined as the gauge-invariant probability to find exactly $n$ electrons outside a predefined sphere $r_0$. Methods for computing these yields are thoroughly discussed. We will denote them by $\mathcal{Y}^{(n)}$. Alternatively, one can conveniently measure the fraction of electrons detached from the atom or molecule by calculating the expectation value of $\hat{\Theta}$ (defined in Eq.~\eqref{Def:Theta}), whose explicit form in spin-adapted formalism is
\begin{equation}
\mathcal{N}=\sum_{ij}\rho_{ij}\Theta_{ij}.
\label{CalcYield}
\end{equation}
It is important to realize that $\mathcal{Y}^{(n)}$ is different from $\mathcal{N}$ in general cases, as the sum of the former never exceeds unity, while the latter can be as large as the number of electrons:
\begin{subequations}
\begin{align}
1&=\sum_{n=0}^{N} \mathcal{Y}^{(n)}, \\
\mathcal{N}&=  \sum_{n=1}^{N} n\mathcal{Y}^{(n)}.
\end{align}
\end{subequations}
However, under the conditions dominated by single ionization, we can safely approximate $\mathcal{Y}^{(n)}\approx0$ for $n>1$, so that $\mathcal{N}\approx\mathcal{Y}^{(1)}\approx1-\mathcal{Y}^{(0)}$.
Analogous to the procedure used for calculating the PMD, we define the channel-resolved photoelectron number collecting the contributions from the $|\gamma\alpha\rangle$ and $|\gamma\beta\rangle$ channels as
\begin{equation}
\mathcal{N}_{\gamma}=2\sum_{ij\in\mathcal{A}} D^*_{\gamma i}D_{\gamma j} \Theta_{ij},
\label{CRY}
\end{equation}
which satisfies
\begin{subequations}\begin{align}
\mathcal{N}_{\gamma}&=\int d^3\bm{k}\mathcal{P}_\gamma(\bm{k}),\\
\mathcal{N}&=\sum_{\gamma}\mathcal{N}_{\gamma}.
\end{align}\end{subequations}
As verified in subsequent sections, under single-ionization dominance, $\mathcal{N}_{\gamma}$ corresponds to the population of the ion in the state $|\gamma\rangle$, which makes it particularly useful for calculating the partial photoionization cross sections.
\subsection{Photoion's Density Matrix}
In spin-adapted formalism, the elements of the iRDM are equal for the two spin subspaces,
\begin{subequations}
\begin{align}
\rho^{(\rmi)}(\gamma\alpha,\gamma'\alpha)&=\rho^{(\rmi)}(\gamma\beta,\gamma'\beta)\equiv\frac{1}{2}\tilde{\rho}^{(\rmi)}(\gamma,\gamma'),   \\
\rho^{(\rmi)}(\gamma\alpha,\gamma'\beta)&=\rho^{(\rmi)}(\gamma\beta,\gamma'\alpha)=0,
\end{align}
\end{subequations}
where we have defined the spin-adapted iRDM $\tilde{\rho}^{(\rmi)}$ normalized to $\mathrm{Tr}\tilde{\rho}^{(\rmi)}=1$.
The spectral decomposition of the ion density operator is simplified into
\begin{equation}
\hat{\rho}^{(\rmi)} = \sum_{\gamma\gamma'} \frac{1}{2}\tilde{\rho}^{(\rmi)}(\gamma,\gamma')\sum_\sigma|\gamma\sigma\rangle\langle\gamma'\sigma|,
\end{equation}
where the matrix elements corresponding to Eqs.~\eqref{iRDM} and \eqref{iRDM2} are calculated by
\begin{equation}
\label{Calc:RDMI}
\begin{aligned}
\tilde{\rho}^{(\rmi)}(\gamma,\gamma') 
&= \frac{2}{\mathcal{N}} \int d^3\bm{k}\mathcal{Q}^*_{\gamma'}(\bm{k})\mathcal{Q}_{\gamma}(\bm{k})\\
&= \frac{2}{\mathcal{N}} \sum_{ij\in\mathcal{A}} D^*_{\gamma' i}D_{\gamma j}\Theta_{ij}.
\end{aligned}
\end{equation}
One may immediately recognize from Eq.~\eqref{CRY} that the diagonal entry $\tilde{\rho}^{(\rmi)}(\gamma,\gamma)=\mathcal{N}_\gamma/\mathcal{N}$ is the branching ratio, ignoring the subsequent nuclear dynamics. 
The off-diagonal entry $\tilde{\rho}^{(\rmi)}(\gamma,\gamma')$ is a measure of quantum coherence between $\gamma$ and $\gamma'$. 
Non-zero off-diagonal entries in iRDM may induce oscillations at frequencies of $\omega_{\gamma\gamma'}=|E_{\gamma}-E_{\gamma'}|=|\mathcal{E}_{\gamma'}-\mathcal{E}_{\gamma}|$ in the ion charge density
\begin{equation}
\label{Calc:Idens}
\begin{aligned}
\varrho^{(\rmi)}(\bm{r})
&=\sum_{\gamma\gamma'}\frac{1}{2}\tilde{\rho}^{(\rmi)}(\gamma,\gamma')\sum_{\sigma\sigma'}\langle\gamma'\sigma'|\hat{a}^\dag_{\bm{r}\sigma}\hat{a}_{\bm{r}\sigma}|\gamma\sigma'\rangle \\
&=\frac{1}{\mathcal{N}}\sum_{ijkl\in\mathcal{A}}\rho_{ijkl}\Theta_{il}\phi_j^*(\bm{r})\phi_k(\bm{r}),
\end{aligned}
\end{equation}
leading to the migration and localization of charge~\cite{rodriguez2024core,calegari2014ultrafast,kraus2015measurement,he2022filming}. In the next section, we will present a real-time simulation of this process, considering fully the detached electron and the residual ion. To visualize it, we calculate the hole density by the difference between the charge density of the neutral species and the ion:
\begin{equation}
\begin{aligned}
\mathcal{\rho}^{(\rm h)}(\bm{r})
&=\mathcal{\rho}^{(\rm n)}(\bm{r})-\mathcal{\rho}^{(\rmi)}(\bm{r}) \\
&\approx \sum_{ij\in\mathcal{A}}\bigg(\rho_{ij}-\frac{1}{\mathcal{N}}\sum_{kl\in\mathcal{A}}\rho_{ijkl}\Theta_{il}\bigg)\phi_j^*(\bm{r})\phi_k(\bm{r}).
\end{aligned}
\end{equation}
Here, we have replaced the charge density of the neutral species with the one of the full system, which is only valid in the weak-field regime. Nevertheless, the integral of the hole density over the entire space still equals exactly unity.\\
\section{Results and Discussion}\label{Results}
\subsection{Ground State}
\begin{figure*}[!htb]
\includegraphics[width=1.0\textwidth]{\detokenize{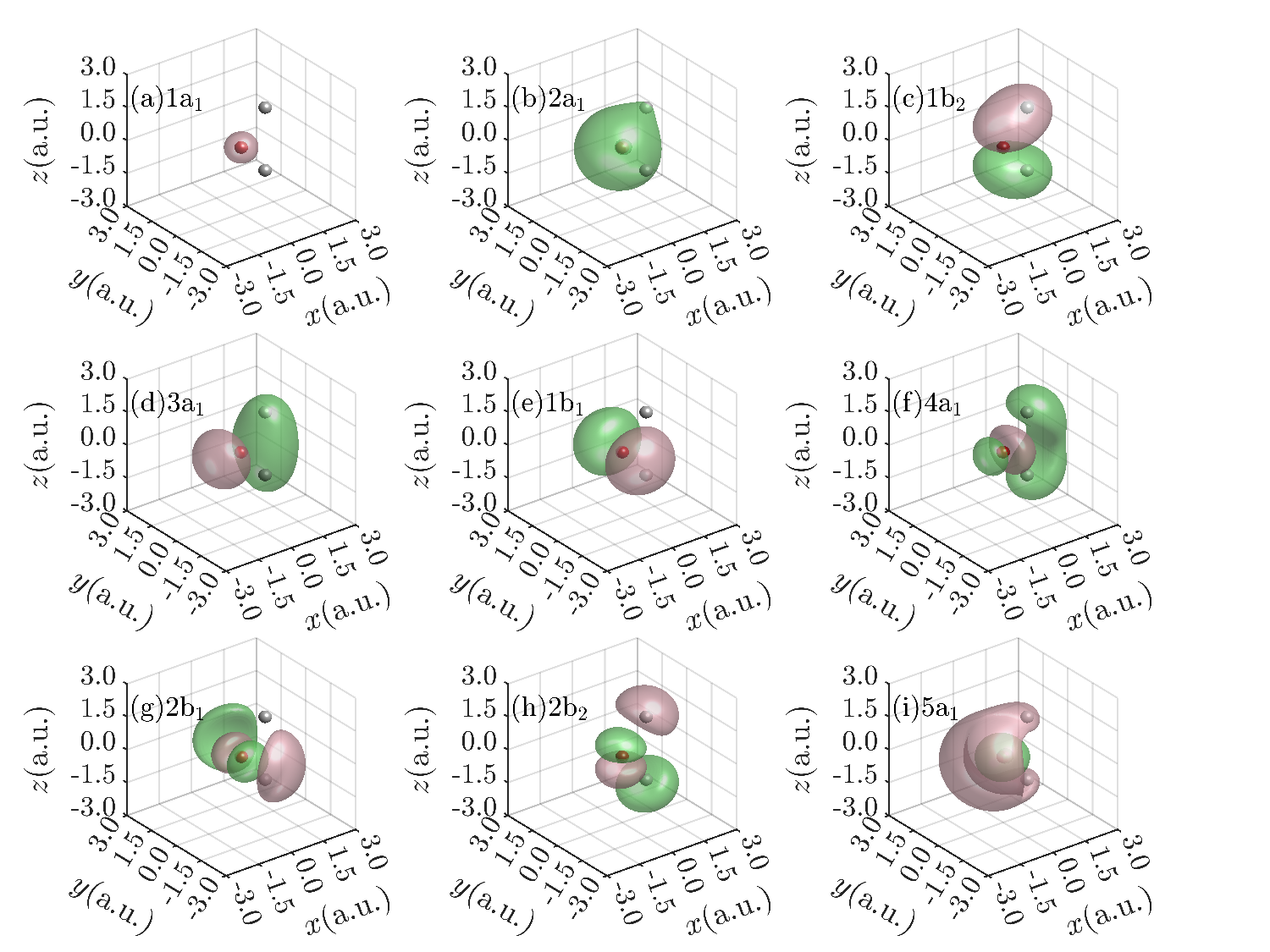}}
\caption{(a)-(i)~The isosurfaces at $\phi_i(\bm{r})=\pm 0.1$ (green for $+$ and red for $-$, colorful online) of $\rm H_2O$ natural molecular orbitals. Data are calculated with imaginary time propagation at (10e,9o). 
The atomic centers for O and H are indicated by small red and grey balls, respectively.}
\label{FigGS}
\end{figure*}

At equilibrium geometry, the $\mathrm{H_2O}$ molecule has $\mathrm{C_{2v}}$ symmetry. In this work, the equilibrium structure is characterized by an O-H bond length of $R_0=1.811$~a.u., and an O-H bond angle fixed at the experimental value of $104.45^{\circ}$. To optimize the convergence of the one-center expansion defined by Eq.~\eqref{monocentricEXP}, we place the oxygen atom at the origin and the two hydrogen atoms at $(x,y,z)=(R_0\cos\Theta_0,0,\pm R_0\sin\Theta_0)$, respectively. In this arrangement, the $x$-$z$ plane is the molecular plane, while both the $x$-$y$ and the $x$-$z$ planes are mirror planes, as shown in Fig.~\ref{FigGS}. The computational efficiency improves significantly for time-dependent calculations when the polarization vector of the laser field satisfies $\bm{\epsilon}\cdot\bm{e}_z=0$, since the reflection symmetry about the $x$-$y$ plane can be exploited.

\begin{table}[!hbpt]
\begin{tabular}{|c|c|c|c|c|}
\hline
CAS		& $-E_{\text{e}}$ & $\mathcal{E}_{X}$ & $\mathcal{E}_{A}$ & $\mathcal{E}_{B}$\\
\hline
\text{(10e,5o)}	&85.2500\{15\}	&	0.5103\{04\}	& 0.5846\{49\}	& 0.7166\{70\}\\
\text{(10e,6o)}	&85.2664\{80\}	& 0.5114\{15\}	& 0.5838\{41\}	& 0.7179\{83\}\\
\text{(10e,7o)} 	&85.3038\{54\}	&	0.5126\{28\}	& 0.5857\{60\}	& 0.7173\{79\}\\
\text{(10e,8o)}	&85.3448\{64\}	& 0.4636\{37\}	& 0.5858\{61\}	& 0.7145\{50\}\\
\text{(10e,9o)}	&85.3820\{35\}	&	0.4664\{65\}	& 0.5504\{06\}	& 0.7181\{86\}\\
\text{(8e,5o)}		&85.2663\{77\}	& 0.5114\{15\}	& 0.5838\{41\}	& 0.7179\{83\}\\
\text{(8e,7o)}		&85.3444\{58\}	& 0.4632\{33\}	& 0.5855\{58\}	& 0.7144\{49\}\\
\text{(8e,8o)}		&85.3809\{25\}	& 0.4657\{58\}	& 0.5496\{98\}	& 0.7181\{86\}\\
Ref.~\cite{brundle1968high}	& N/A 						& 0.4634 			& 0.5413 			& 0.6817\\
Ref.~\cite{potts1972photoelectron} & N/A 	  & 0.4638				& 0.5417				& 0.6802\\
\hline
\end{tabular}
\caption{The ground state energy ($E_e$) and the vertical ionization energy ($\mathcal{E}_{X,A,B}$) of $\rm H_2O$ molecule at equilibrium geometry by the present work (in atomic unit).
The deviation in the last place are given parenthetically, where the reference values are computed with the basis set aug-cc-PV6Z\cite{peterson1994benchmark,wilson1996gaussian}. The last two rows are experimental values via photoelectron spectroscopy.}
\label{GS-H2O}
\end{table}

We validate our numerical approach by comparing the calculated ground state energy ($E_0$, nuclear repulsion energy excluded), and vertical ionization energies ($\mathcal{E}$) with reference values obtained using PySCF~\cite{sun2018pyscf}. ITP are adopted. Specifically, the vertical ionization energies, $\mathcal{E}_{X}$, $\mathcal{E}_{A}$, and $\mathcal{E}_{B}$, defined as the energies required to remove an electron from $\rm 1b_1$, $\rm 3a_1$, and $\rm 1b_2$, respectively, are calculated by Eq.~\eqref{spin-adapted-EKT}. The corresponding ion states are conventionally denoted as $X(\mathrm{{1}^{2}B_1})$, $A(\mathrm{{1}^{2}A_1})$ and $B(\mathrm{{1}^{2}B_2})$, whose dominant single-vacancy states are $\rm 1b^{-1}_1$, $\rm 3a^{-1}_1$ and $\rm 1b^{-1}_2$, respectively. We employ a large basis with parameters $N_m=21$, $N_l=22$, $m_0=-10$, $\lambda_{\max}=15$, and $\mu_{\max}=7$, to ensure convergence approaching the complete basis limit. The box size is $r_{\max}=36$~a.u.. We use the conventional notation (Ne,Mo) for CAS, where N electrons occupy M spatial orbitals, while the remaining electrons reside in doubly-occupied frozen-core orbitals. The initial guesses of the orbital functions are obtained either from the atomic orbitals of oxygen or from a previous run, which are then fitted into the desired irreducible representation of the orbitals. The (10e,5o) calculation is initialized according to the Hartree-Fock configuration of the ground state $(\mathrm{{}^1A_1})$, $\rm (1a_1)^2 (2a_1)^2 (1b_2)^2 (3a_1)^2 (1b_1)^2$. Subsequent to (10e,5o), we expand the CAS size upward to (10e,9o), with known symmetries ($\rm 4a_1$, $\rm 2b_2$, $\rm 2b_1$, and $\rm 5a_1$) guiding orbital assignments. The results are listed in Tab.~\ref{GS-H2O}. Experimental values are also presented for comparison when available~\cite{brundle1968high,potts1972photoelectron}.

In all calculations, both the ground-state and vertical ionization energies agree well with the reference value obtained using PySCF with an aug-cc-PV6Z basis set~\cite{peterson1994benchmark,wilson1996gaussian}. The differences are confined to the last two digits, underscoring the accuracy of the one-center expansion technique.
In addition, we observe that $\mathcal{E}_{X}$ is significantly improved with the inclusion of $\rm 2b_1$.
Adding $\rm 5a_1$ orbital greatly improves $\mathcal{E}_{A}$ as well. However, $\mathcal{E}_{B}$ still differs from the experimental value by approximately 1~eV for (10e,9o), and adding more orbitals only shows a small improvement. 
We thus stop at (10e,9o) due to computational resource constraints.
A comparison between (8e,8o) and (10e,9o) suggests that freezing $\rm 1a_1$ is a good approximation for the study of dynamics involving only valence electrons.
For visualization, the isosurface plots of the involved molecular natural orbitals at (10e,9o) are shown in Fig.~\ref{FigGS}.\\
\subsection{Photoionization Cross Sections}\label{tXsec}
In this section, the orientation-averaged total and partial photoionization cross sections of $\mathrm{H_2O}$ are calculated and compared with established theoretical values~\cite{diercksen1982theoretical,cacelli1992molecular,ruberti2013total,Bruno2016,modak2020probing,Benda2020,fernandez2023photoionization} and experimental data~\cite{tan1978absolute,haddad1986total,chan1993electronic}. In this work, we focus on scenarios with photon energies larger than $\mathcal{E}_{B}$.
Below the third ionization threshold ($\mathcal{E}_{B}$), numerous prominent doubly-excited auto-ionizing states occur~\cite{fernandez2023photoionization}, which require a very large orbital space for an accurate description. Besides, the co-existence of nuclear motion and autoionization further complicates the theoretical treatment~\cite{palacios2015theoretical}, exceeding the capability of our current approach.

By irradiating the body-fixed molecule with a linearly polarized pulse and calculating the probability $\mathcal{Y}$ for single-photon absorption, the photoionization cross sections are determined. The cross section $\sigma$ is related to $\mathcal{Y}$ via~\cite{colgan2003time}
\begin{equation}
\sigma=\frac{8\pi}{\omega c A_{\max}^2} \frac{\mathcal{Y}}{T_\mathrm{eff}},
\label{CalcXsec}
\end{equation}
where $\omega=2\pi/T$ is the central frequency, $c\approx137.036$ is the speed of light, $A_{\max}$ is the amplitude of the vector potential, $T_{\rm eff}$ is the effective pulse duration. Within the dipole approximation, the orientation average for single-photon absorption is simplified into 
\begin{equation}
\mathcal{Y}=\frac{1}{3}(\mathcal{Y}_x+\mathcal{Y}_y+\mathcal{Y}_z),
\end{equation} 
where $\mathcal{Y}_{x,y,z}$ represents the ionization probabilities for laser polarization vector $\bm{\epsilon}$ aligned with $\bm{e}_{x,y,z}$, respectively. 

\begin{figure}[!htb]
\includegraphics[width=0.5\textwidth]{\detokenize{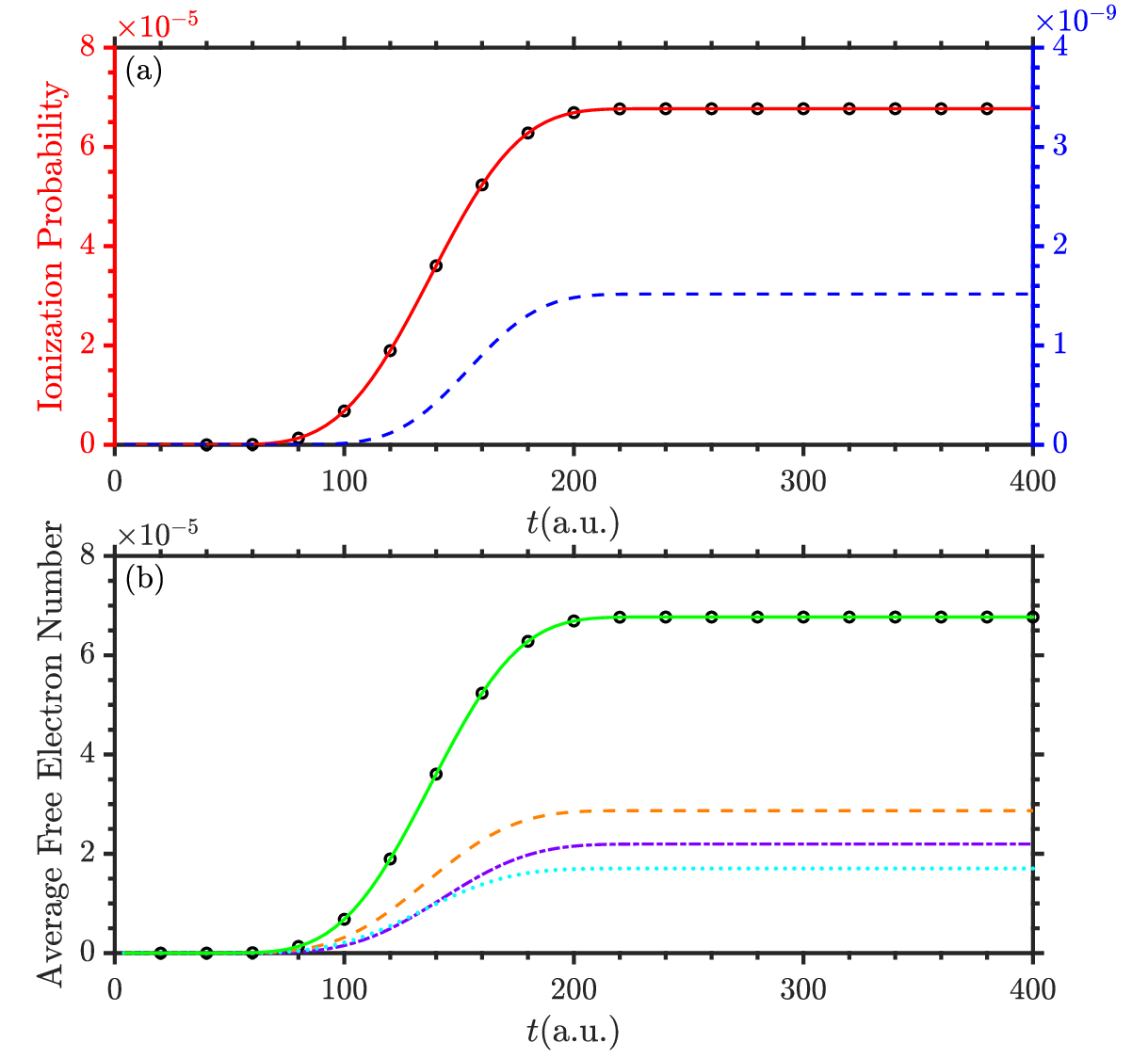}}
\caption{(a) The time-dependent probabilities of a body-fixed $\rm H_2O$ molecule in (8e,8o), driven by a 32-cycle sine-square pulse with a central frequency of $\omega=1$~a.u.\,and $\bm{\epsilon}=\bm{e}_x$. The geometry is found in Fig.~\ref{FigGS}. The black circles, red solid line and blue dashed line indicate $1-\mathcal{Y}^{(0)}$, $\mathcal{Y}^{(1)}$ and $\mathcal{Y}^{(2)}$, respectively. (b) The time-dependent channel-resolved average free electron numbers in the same calculation. The orange dashed line, purple dash-dot line and cyan dot line indicate the one for $B({1}^2B_2)$, $A({1}^2A_1)$ and $X({1}^2B_1)$, respectively. The green solid line is their sum.}
\label{FigYield}
\end{figure}

In practical calculations, we use an $M$-cycle sin-square pulse, where $T_{\rm eff}=3MT/8$. We choose $M=32$ to approach the long-pulse limit and a small vector potential amplitude $A_{\text{max}}=5\times 10^{-4}$~a.u. to stay within the perturbative regime. For all calculations in this section, we use a small expansion where $r_{\max}=120$~a.u., $m_0=-8$, $N_m=17$, $N_l=16$ and $N_b=256$. The time step is kept at $0.004$~a.u. always. An exponential absorber is applied to the region $r\ge56~$a.u. at each step, which is sufficient to suppress boundary reflections. To obtain the single-photon single-ionization cross sections, we adopt the approximation $\mathcal{Y}\equiv \mathcal{Y}^{(1)}\approx1-\mathcal{Y}^{(0)}\approx\mathcal{N}$, allowing the application of  Eq.~\eqref{CalcYield}.
This approximation was validated by comparing $\mathcal{Y}^{(i)}$, $i=0,1,2$ with $\mathcal{N}$ as a function of time in a test calculation at $\omega=1.0$~a.u. and polarization along $\bm{e}_x$, as shown in Fig.~\ref{FigYield}(a). Photoelectron wavepackets surpassing $r_0=32$~a.u. are considered fragments. We find that $\mathcal{Y}^{(2)}$ are $10^{-4}-10^{-5}$ times smaller than $\mathcal{Y}^{(1)}$, and $1-\mathcal{Y}^{(0)}$ is in good agreement with $\mathcal{Y}^{(1)}$. The average number of electrons in the exterior region $(\mathcal{N})$ is given by summing over all $\mathcal{N}_\gamma$, as shown in Fig.~\ref{FigYield}(b), and also agrees well with $1-\mathcal{Y}^{(0)}$. These comparisons reveal negligible contributions from multi-ionization processes, validating our approximation. 
The partial cross sections for the ionization pathways
\begin{subequations}
\begin{align}
\mathrm{H_2O}+\omega&\longrightarrow \mathrm{H_2O^+}(X) + \mathrm{e^-},\\
\mathrm{H_2O}+\omega&\longrightarrow \mathrm{H_2O^+}(A) + \mathrm{e^-},\\
\mathrm{H_2O}+\omega&\longrightarrow \mathrm{H_2O^+}(B) + \mathrm{e^-},
\end{align}
\end{subequations}
are calculated and denoted as $\mathcal{N}_{X}$, $\mathcal{N}_{A}$ and $\mathcal{N}_{B}$, with the total cross section $\mathcal{N}=\mathcal{N}_{X}+\mathcal{N}_{A}+\mathcal{N}_{B}$.

\begin{figure}[!htb]
\includegraphics[width=0.5\textwidth]{\detokenize{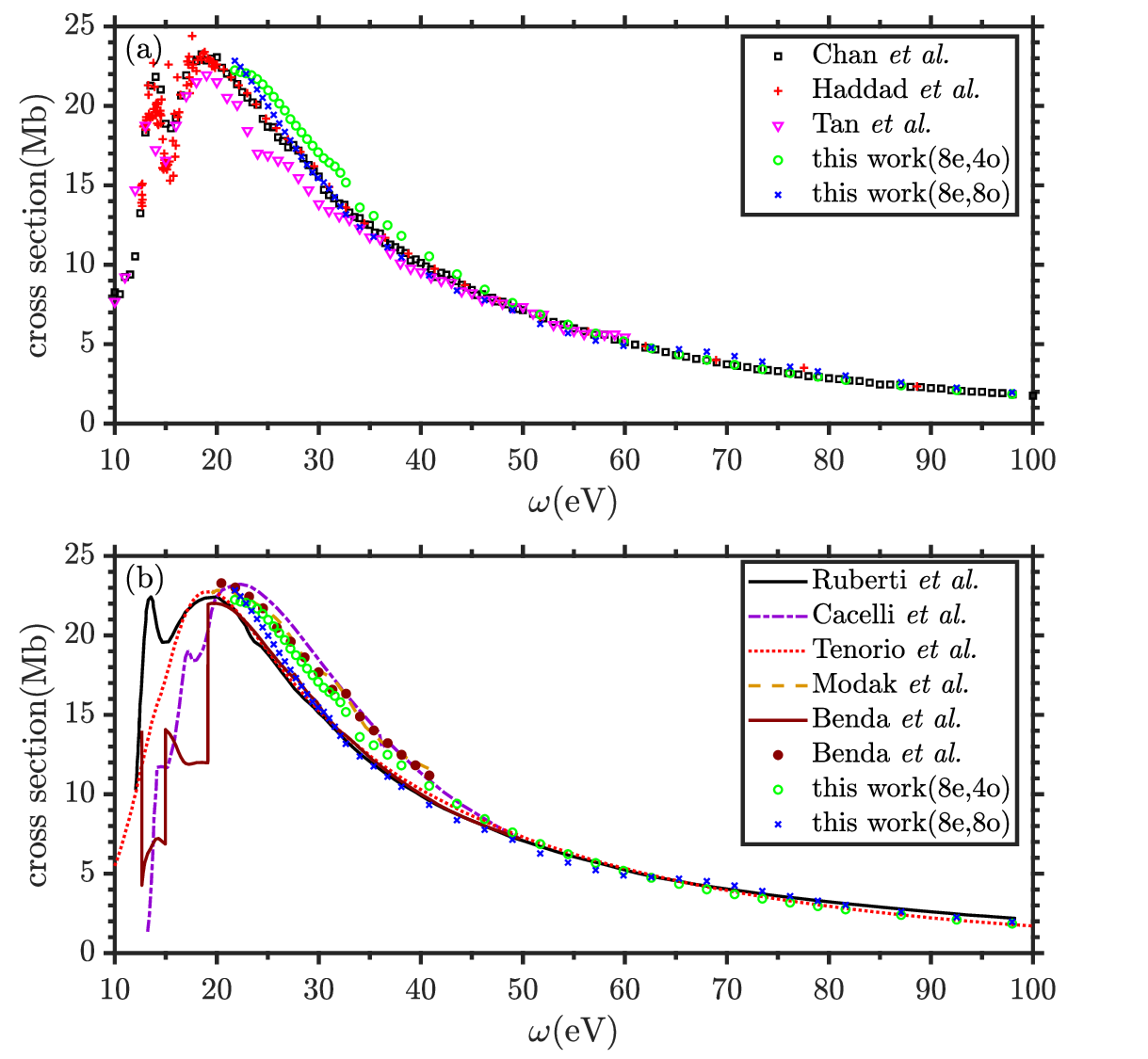}}
\caption{The total photoabsorption cross section of orientation averaged $\rm H_2O$ obtained in this work at (8e,4o) (green empty circles) and at (8e,8o) (blue crosses), (a) compared with earlier experiments by Tan \textit{et al.}\cite{tan1978absolute} (magenta empty triangles), Haddad and Samson\cite{haddad1986total} (red plus signs), Chan \textit{et al.}\cite{chan1993electronic} (black empty squares), or (b) compared with earlier theories by Ruberti \textit{et al.}\cite{ruberti2013total} (black solid line), Tenorio \textit{et al.}\cite{Bruno2016} (red dotted line), Benda \textit{et al.}\cite{Benda2020} (brown solid line for R-matrix and brown solid circles for RMT), Modak and Antony\cite{modak2020probing} (yellow dashed line), Cacelli \textit{et al.}\cite{cacelli1992molecular} (purple dot-dashes line). Reference data are extracted with digitization tools.}
\label{FigXSEC}
\end{figure}

Our calculated total photoionization cross sections are compared with experimental~\cite{tan1978absolute,haddad1986total,chan1993electronic} and theoretical references~\cite{ruberti2013total,Bruno2016,Benda2020,modak2020probing}, as shown in Figs.~\ref{FigXSEC}(a) and~\ref{FigXSEC}(b), respectively. The gauge invariance has been checked by comparing the numerical value of the cross sections in the length gauge and the velocity gauge.
Notably, our (8e,8o) calculations yield improved agreement with experimental data between 20~eV and 40~eV, whereas the result at (8e,4o) level overestimates the cross sections by 2–4 Mb.
For higher photon energies, discrepancies emerge from inaccurate energies of resonant states due to the limited CAS size, though differences remain minor.
For photon energies near the threshold of $B$ state, (8e,8o) results systematically overestimate the experimental values by 2-4 Mb. This can be foreseen by the difference in the ionization threshold into $B$ state in theory (19.5~eV) and experiment (18.5~eV), suggesting that the third channel has not fully converged. Note that we do not manually shift the curve to match the thresholds.
As shown in Fig.~\ref{FigXSEC}(b), our results align closely with recent accurate theoretical calculations in general. In particular, the ADC(2) calculation by Ruberti \textit{et al.}~\cite{ruberti2013total}, the coupled-cluster calculation by Tenorio \textit{et al.}~\cite{Bruno2016} and the R-matrix calculation performed with UKRmol+ with a large CAS by Benda \textit{et al.}~\cite{Benda2020} produce the closest result to the experimental values.  
Due to the insufficient description of electron correlation, the RPA result by Cacelli \textit{et al.}~\cite{cacelli1992molecular} based on a Hartree-Fock calculation, the R-matrix result by Modak and Antony~\cite{modak2020probing} which describes the cation states at the Hartree-Fock level, the RMT result by Benda \textit{et al.}\cite{Benda2020} which describes the cation states at (7e,5o) and our (8e,4o) result at TDHF level systematically higher estimate the cross section in the interval between 20~eV to 40~eV. While our (8e,8o) result better matches the more accurate time-independent theories where more electron correlation effects are involved. We also mention the work by Fernandez \textit{et al.}\cite{fernandez2023photoionization}, where XChem is applied to resolve the resonance structures below $\mathcal{E}_{B}$ and not shown here.

\begin{figure}[!htb]
\includegraphics[width=0.5\textwidth]{\detokenize{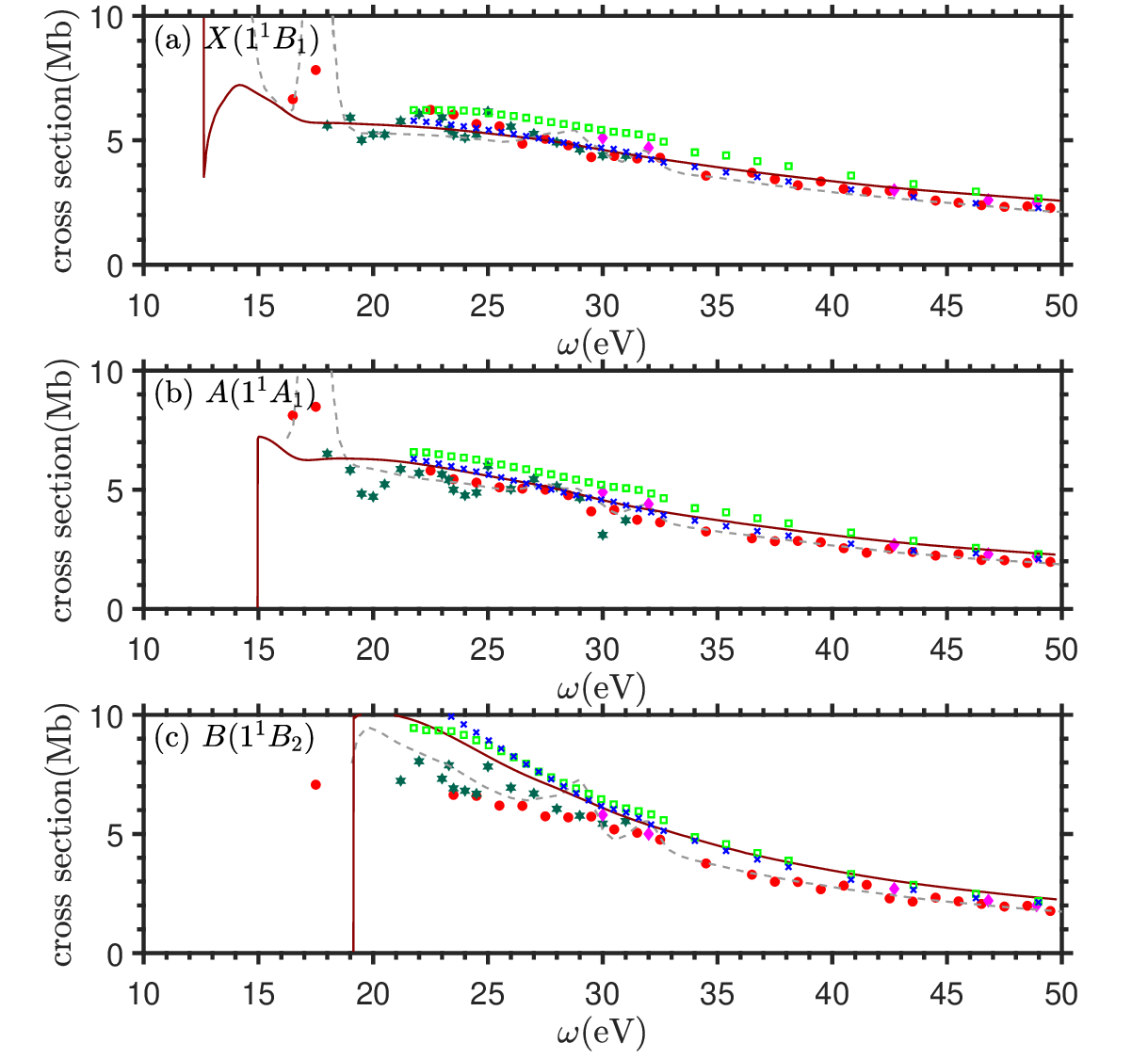}}
\caption{The partial photoabsorption cross section of orientation averaged $\rm H_2O$ obtained in this work at (8e,4o) (green empty circles) and at (8e,8o) (blue crosses), compared with earlier experimental measurements by Tan \textit{et al.}\cite{tan1978absolute} (red solid circles), Truesdale \textit{et al.}\cite{truesdale1982photoelectron} (dark green solid hexagrams), Banna \textit{et al.}\cite{banna1986photoelectron} (magenta solid diamonds), and the theories by Stener \textit{et al.}\cite{stener2002time}(gray dashed line), Benda \textit{et al.}\cite{Benda2020} (brown solid line). (a) $X$ (b) $A$ (c) $B$. Reference data are extracted with digitization tools.}
\label{FigXSECCR}
\end{figure}

We compare the resultant partial photoionization cross sections with the reference works in Fig.~\ref{FigXSECCR}, including the experimental measurements by Tan \textit{et al.}\cite{tan1978absolute}, Truesdale \textit{et al.}\cite{truesdale1982photoelectron}, Banna \textit{et al.}\cite{banna1986photoelectron}, and the theories by Stener \textit{et al.}\cite{stener2002time}, Benda \textit{et al.}\cite{Benda2020}. For both $X$ and $A$ channels, the calculation (8e,8o) agrees well with the experimental measurements in the interval between 20~eV to 40~eV as well as the accurate R-matrix results with a larger CAS by Benda \textit{et al.}\cite{Benda2020}, as shown in Fig.~\ref{FigXSECCR}(a)(b). The systematic higher estimation in the (8e,4o) results, as well as the R-matrix results by Modak and Antony~\cite{modak2020probing} and RPA by Cacelli \textit{et al.}\cite{cacelli1992molecular}(not shown here) suggests again that the ion states are poorly described at the Hartree-Fock level. In Fig.~\ref{FigXSECCR}(c), we find that (8e,8o) still slightly higher estimate the partial cross section near 25~eV, compared with the accurate R-matrix results. The size of CAS could be insufficient for an accurate description of the $B$ state in our calculation. We also mention that most recent theories in Fig.~\ref{FigXSECCR}(c), except the earlier TD-DFT results by Stener~\cite{stener2002time}, produce prominently higher values than all available experiments. However, such a discrepancy does not occur when these theories are compared with the most recent measurements by Chan \textit{ et al.}\cite{chan1993electronic} for the total cross section in Fig.~\ref{FigXSEC}.

At this moment, we have illustrated the power of our channel-resolved formalism via a rigid comparison of the total and partial photoionization cross sections with benchmark values. With the channel-resolved formalism, partial cross sections can be provided without explicitly logging the PMD, which may save enormous computational resources for similar applications.
Furthermore, it reminds us that the full \textit{ab initio} wavefunction that is conventionally overlooked as a black box might embed uncovered aspects than it seems.\\
\subsection{Control of Attosecond Coherence}
In this section, we apply our channel-resolved formalism to control the quantum coherence by manipulating the polarization and the duration of the pulse. We employ a pair of phase-delayed attosecond pulses which reads
\begin{subequations}
\begin{align}
	A_x(t) &=      A_{\max} f(t,M)\sin(\omega t), \\
	A_z(t) &= \eta A_{\max} f(t,M)\sin(\omega t+\chi).
\end{align}
\end{subequations}
Here, $f(t,M)$ is the envelope, and $M$, $\chi$, $\eta$ control the number of cycles, phase delay, and relative strength of the two attosecond pulses. We fix $\omega=1.4$~a.u.. The basis parameters $m_0=-12$, $N_m=25$, $N_l=18$, $N_b=400$, and $r_{\max}=240$~a.u. are used in all calculations in this section. The time step is kept at $0.005$~a.u., and an exponential absorber is applied to the region $r\ge 200$~a.u. at each step. 

As the laser is polarized in the molecular plane, $\tilde{\rho}^{(\rmi)}(X,B)$ and $\tilde{\rho}^{(\rmi)}(X,A)$ are strictly zero due to the symmetry. To understand it, we recall the table of dipole selection rules of the point group $\mathrm{C_{2v}}$
\begin{equation}
\begin{matrix}
            &\mathrm{a_1} & \mathrm{b_1} & \mathrm{a_2} & \mathrm{b_2}\\
\mathrm{a_1}&  x  &  y  &      & z\\
\mathrm{b_1}&  y  &  x  &  z   & \\
\mathrm{a_2}&     &  z  &  x   & y\\
\mathrm{b_2}&  z  &     &  y   & x\\
\end{matrix}
\end{equation}
and that $X$, $A$ and $B$ are approximately formed by removing an electron from $\mathrm{1b_1}$, $\mathrm{3a_1}$ and $\mathrm{1b_2}$, respectively.
In the absence of $A_y(t)$, the laser only couples $\mathrm{b_1}$ with $\mathrm{a_2}$ and $\mathrm{a_1}$ with $\mathrm{b_2}$, hence the integral over $\mathcal{Q}^*_X(\bm{k})\mathcal{Q}_{A}(\bm{k})$ and $\mathcal{Q}^*_X(\bm{k})\mathcal{Q}_{B}(\bm{k})$ is strictly zero, whereas the one over $\mathcal{Q}_{A}^*(\bm{k})\mathcal{Q}_{B}(\bm{k})$ could be non-zero.
Therefore, we focus on $A$ and $B$, using $\tilde{\rho}^{(\rmi)}(A,B)$ as a measure of the quantum coherence between them.

\begin{figure}[!htb]
\includegraphics[width=0.5\textwidth]{\detokenize{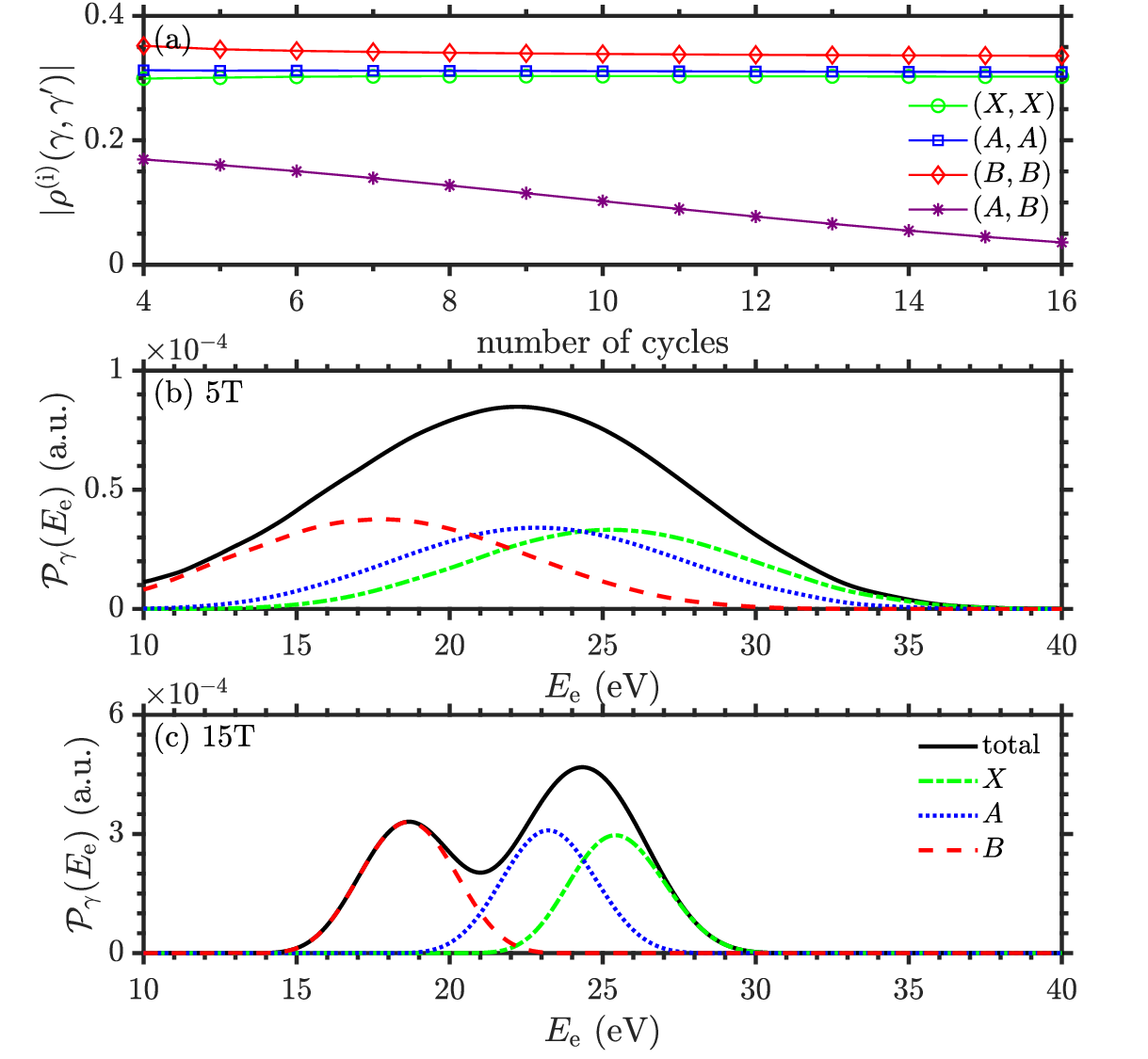}}
\caption{(see text for detailed laser parameters) 
(a) The magnitude of $\tilde{\rho}^{(\rmi)}(X,X)$ (green circles), $\tilde{\rho}^{(\rmi)}(A,A)$ (blue squares), $\tilde{\rho}^{(\rmi)}(B,B)$ (red lozenges), $\tilde{\rho}^{(\rmi)}(B,A)$ (green circles) of a body-fixed water molecule irradiated by a circularly polarized attosecond pulse with a varying duration $M$. (b) The total photoelectron energy distributions (black solid line) and the one in channel $X$ (green dot-dashed line), $A$ (blue dotted line) and $B$ (red dashed line). (c) The same to (b) but at $M=15$.}
\label{FigCohereNcycle}
\end{figure}

In the first scenario, we set $\eta=1$, $A_{\max}=10^{-3}$~a.u., and $\chi=\pi/2$ to generate a circularly polarized pulse. We then vary $M$ to examine the behavior of iRDM. As shown in Fig.~\ref{FigCohereNcycle}(a), the diagonal entries remain relatively strong across all values of $M$, and their changes are almost invisible. However, the magnitude of off-diagonal entry $\tilde{\rho}^{(\rmi)}(A,B)$ decreases rapidly from one-half of $\tilde{\rho}^{(\rmi)}(A,A)$ and $\tilde{\rho}^{(\rmi)}(B,B)$ to an extent below 0.05. This phenomenon is expected to occur because the energetic overlap between the partial PMDs is smaller for longer pulses, such that the off-diagonal entry also decreases according to Eq.~\eqref{Calc:RDMI}. To show it clearly, we calculate the total and partial photoelectron energy distributions (PEDs) by integrating out the angular variables in the PMDs obtained via the projection methods in Sec.~\ref{CRPMD1}. The resulting PEDs at $M=5,15$ are given in Fig.~\ref{FigCohereNcycle}(b)(c), respectively. The peaks in channel $A$ and $B$ are very diffusive at $M=5$ and share a large overlap area as in Fig.~\ref{FigCohereNcycle}(b). In this case, it is impractical to distinguish $A$ from $B$ through the total PED. In contrast, they are separable by the naked eye at $M=15$ and the overlap area shrinks, as in Fig.~\ref{FigCohereNcycle}(c), giving rise to the observed phenomena.

\begin{figure}[!htb]
\includegraphics[width=0.5\textwidth]{\detokenize{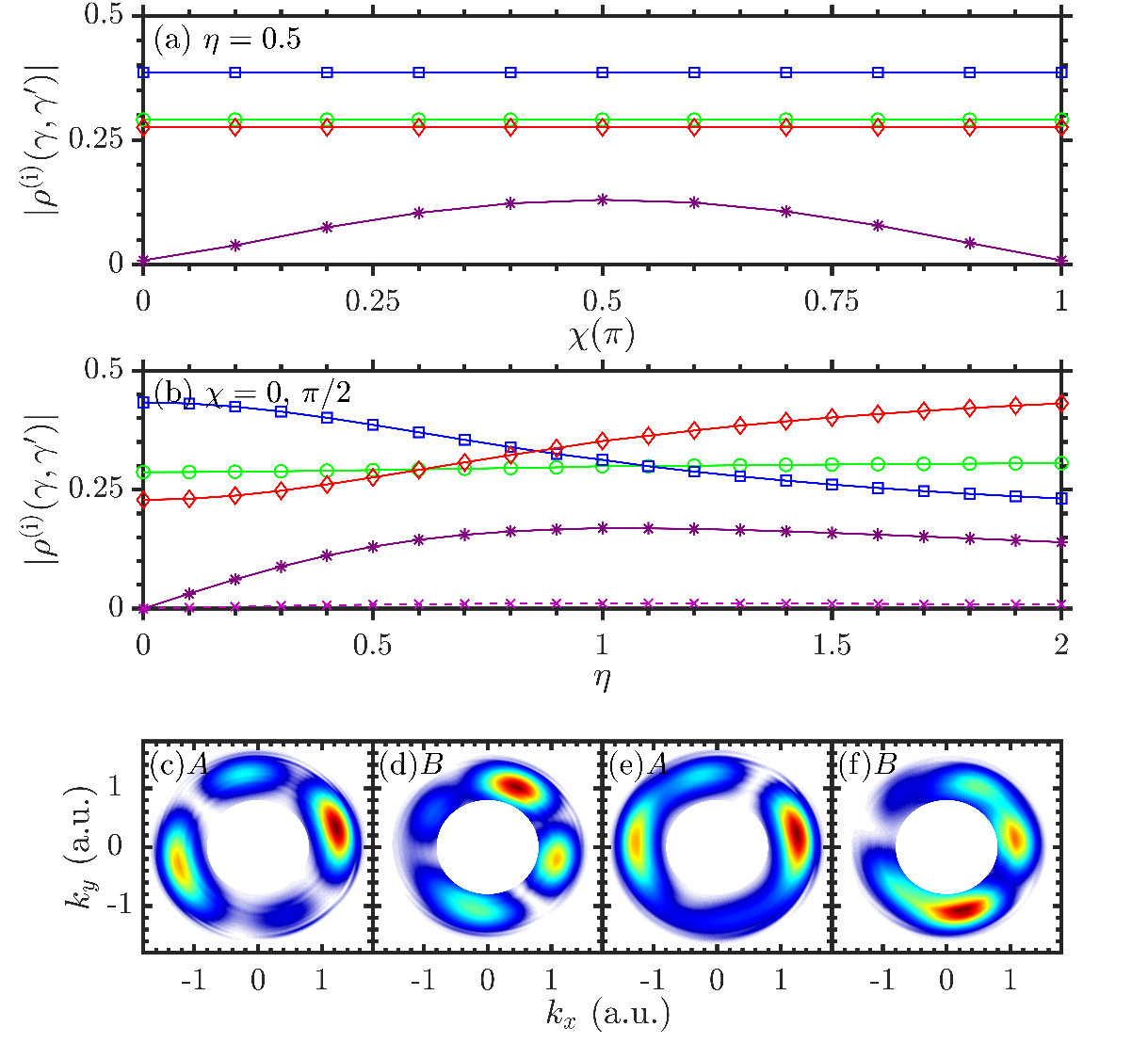}}
\caption{(see text for detailed laser parameters) 
(a) The same to Fig.~\ref{FigCohereNcycle}(a) but the laser field constitutes a pair of orthogonally polarized attosecond pulses with $\eta=0.5$ and varying $\chi$. 
(b) The same to (a) but with $\chi=0,0.5\pi$ and varying $\eta$. The PMDs in $k_x-k_z$ plane between $0.8\leq|\bm{k}|\leq1.8$ (a.u.) are normalized to maximum.
(c)\,Channel $A$, linearly polarized;
(d)\,Channel $B$, linearly polarized;
(e)\,Channel $A$, circularly polarized;
(f)\,Channel $B$, circularly polarized.}
\label{FigCoherePolarization}
\end{figure}

In the second scenario, we set $A_{\max}=10^{-3}$~a.u. and $M=4$. Using varying $\eta$ and $\chi$, we explore how to control the coherence between $A$ and $B$. Due to the perturbative nature of single-photon ionization, the channel-resolved yields do not depend on $\chi$. To understand it, we separate the contributions of $A_s$ $(s=x,z)$ in the final amplitude $\mathcal{Q}_{\gamma}(\bm{k})$ $(\gamma=A,B)$ using the lowest order perturbation theory (LOPT) to write
\begin{equation}
        \mathcal{Q}_{\gamma}(\bm{k})=\mathcal{Q}^{x}_{\gamma}(\bm{k}) + \eta e^{-i\chi} \mathcal{Q}^{z}_{\gamma}(\bm{k}),
\end{equation}
and we define
\begin{equation}
    \mathcal{R}^{ss'}_{\gamma\gamma'} = 
    \int \mathcal{Q}^{s*}_\gamma(\bm{k})\mathcal{Q}^{s'}_{\gamma'}(\bm{k})d^3\bm{k}.
\end{equation}
Due to symmetry, $\mathcal{R}^{zx}_{\gamma\gamma}$ and $\mathcal{R}^{xz}_{\gamma\gamma}$ are strictly zero for any $\gamma$, such that $\mathcal{N}_\gamma=\mathcal{R}^{xx}_{\gamma\gamma}+\eta^2\mathcal{R}^{zz}_{\gamma\gamma}$ is irrelevant to $\chi$. To confirm it, we conduct simulations with a varying $\chi$ but fixed $\eta=0.5$. The results shown in Fig.~\ref{FigCoherePolarization}(a) qualitatively reproduce the prediction from LOPT. Unlike the diagonal entries, the off-diagonal entries are sensitive to the phase-delay between the two pulses, as is evident from its analytic expression
\begin{equation}
    \rho^{(\rmi)}(B,A)=\frac{\eta}{\mathcal{N}^x+\eta^2\mathcal{N}^z}\left(
    e^{-i\chi}\mathcal{R}^{xz}_{AB}+e^{i\chi}\mathcal{R}^{zx}_{AB}\right)
    \label{LOPT-IRDM}
\end{equation}
where $\mathcal{N}^s$ is the sum of $\mathcal{R}^{ss}_{\gamma\gamma}$ over all $\gamma$-s. Eq.~\eqref{LOPT-IRDM} indicates a cosine modulation of $|\tilde{\rho}^{(\rmi)}(B,A)|^2$ with respect to $\chi$ at a period of $\pi$, which is observed in Fig.~\ref{FigCoherePolarization}(a). The maximum locates closely at $0.5\pi$. Generally, for fixed $\eta$, the coherence is maximized or minimized at $\chi=\chi_{2n}$ or $\chi=\chi_{2n+1}$, where
\begin{equation}
    \chi_n =\frac{n\pi+\arg(\mathcal{R}^{xz}_{AB})-\arg(\mathcal{R}^{zx}_{AB})}{2}.
\end{equation}

As another prediction from Eq.~\eqref{LOPT-IRDM}, $A$ and $B$ are maximally coherent when $\eta^2=\mathcal{N}^x/\mathcal{N}^z$ for any given delay $\chi$, which is confirmed in Fig.~\ref{FigCoherePolarization}(b) where the maximum of $|\tilde{\rho}^{(\rmi)}(B,A)|$ is found near $\eta\approx1$ for both $\chi=0$ (linearly polarized) and $\chi=0.5\pi$ (elliptically polarized).
The coherence vanishes when $\eta=0$, which is consistent with the perspective of symmetry. For all $\eta$, large coherence between $A$ and $B$ may not be simply produced by changing the polarization angle of the linearly polarized pulse. The channel-resolved PMDs offer an intuitive picture to explain this property. In Fig.~\ref{FigCoherePolarization}(c) and (d), we show the PMD in $k_x-k_z$ plane in channel $A$ and $B$, respectively, driven by a linearly polarized pulse with $\chi=0$, $\eta=1$. The lobes are staggered, suppressing the overlap integral in Eq.~\eqref{Calc:RDMI}. For the circularly polarized case shown in Fig.~\ref{FigCoherePolarization}(e) and (f) where $\chi=\pi/2$, $\eta=1$, the PMDs in channels $A$ and $B$ overlap significantly near $(k_x,k_z)=(1,0)$, leading to prominent coherence.

\begin{figure}[!htb]
\includegraphics[width=0.5\textwidth]{\detokenize{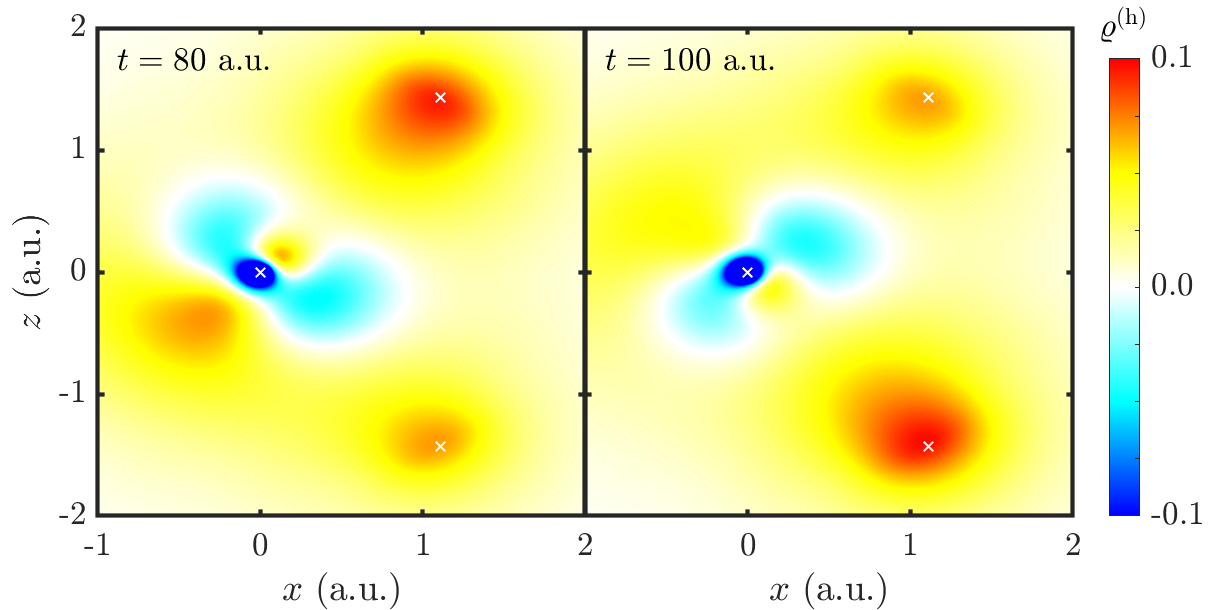}}
\caption{The false color plot of the hole density $\varrho^{(\rm h)}(\bm{r})=\varrho^{(\rm n)}(\bm{r})-\varrho^{(\rmi)}(\bm{r})$ in the molecular plane at $80$~a.u.\,(left panel) and $100$~a.u.\,(right panel) after the laser has damped. White crosses indicate the position of the atoms. Note that the density near the O atom below $-0.1$ is not shown for clarity.}
\label{FigHole}
\end{figure}
As the system has become partially coherent when $\chi=0.5\pi$ and $\eta=1$, it is supposed to find hole dynamics in the simulation, as shown by the difference between the charge density of the neutral molecule and of the ion in Fig.~\ref{FigHole}. At $80$~a.u., the hole is mainly localized near the upper H atom, while at $100$~a.u., it has moved to the lower side. Such an oscillation follows the prediction of Eq.~\eqref{Calc:Idens}, where the period of charge migration is around $2\pi/(\mathcal{E}_B-\mathcal{E}_A)\approx37.3$~a.u..\\

\section{Conclusion}
In summary, we have advanced the theoretical methods for studying attosecond photoionization dynamics in atoms and molecules by introducing the time-dependent hole state formalism into the multi-configurational time-dependent Hartree-Fock framework. Our full-dimensional, one-center expansion numerical implementation enables the simulation of arbitrarily polarized laser fields and multi-center molecular potentials, providing a versatile tool for \textit{ab initio} investigations of ultrafast electron dynamics. The rigorous definition of the time-dependent hole state, derived from a time-domain generalization of the extended Koopmans' theorem, allows accurate extraction of the reduced ion density matrix and channel-resolved observables, capabilities previously unrealized within the general time-dependent multi-configurational self-consistent-field framework.

By applying this approach to the single-photon ionization of the water molecule, we successfully calculated both total and partial photoionization cross sections, achieving excellent agreement with existing experimental and theoretical benchmarks. This validation underscores the reliability of the time-dependent hole state formalism in capturing multichannel ionization processes. Furthermore, we demonstrated the ultrafast control of electron coherence in $\mathrm{H_2O}$ by tailoring the phase delay and polarization of attosecond pulses, highlighting the potential of attosecond spectroscopy to manipulate quantum states in molecular ions.

Our results bridge a critical gap in state-resolved attosecond photoionization studies, offering a generalizable framework for disentangling entangled photofragment dynamics. The compatibility of the methodology with the advanced t-SURFF method paves the way for constructing ion-state-resolved photoelectron momentum distributions in complex systems. While the computational demands of fully converged MCTDHF simulations for large molecules remain non-trivial, the efficiency enhancement strategies in our implementation, such as the gauge-invariant frozen-core approximation and the angular discrete variable representation for accelerating two-electron integrals, extend its applicability to larger molecular targets. Future extensions could incorporate hybrid restricted active
space strategies, such as occupation-restricted multiple active spaces, and non-adiabatic nuclear motion, such as extended MCTDHF, to balance accuracy and scalability, enabling real-time tracking of correlated electron dynamics in ever more complex environments.
\section*{acknowledgment}
This work was supported by National Natural Science Foundation of China (NSFC) (Grant No. 11925405, 12274294) and the Strategic Priority Research Program of Chinese Academy of Sciences (Grant No. XDA25010100). 
This research was also supported in part by a Grant-in-Aid for Scientific Research (Grant No. JP19H00869, JP20H05670, JP22H05025, JP22K18982, and JP24H00427) from the Ministry of Education, Culture, Sports, Science and Technology (MEXT) of Japan. This research was also partially supported by the MEXT
Quantum Leap Flagship Program, (Grant No. JPMXS0118067246)
and JST COI-NEXT, (Grant No. JPMJPF2221).
H.P. gratefully acknowledges support from the RIKEN TRIP initiative (RIKEN Quantum). 
The computations in this paper were run on the Siyuan-1 cluster supported by the Center for High Performance Computing at Shanghai Jiao Tong University.
\section*{Data availability}
The data that support the findings of this work are openly available \cite{data}.

\bibliography{reference}


\appendix
\section{Fundamental Relations}\label{Relations}
We notice that
\begin{subequations}
\begin{align}
	[\hat{H},\hat{a}_\nu]&=-\sum_{\tau\in\mathcal{O}}h_{\nu\tau}\hat{a}_\tau-\sum_{\xi\lambda\tau\in\mathcal{O}}g_{\nu\xi\lambda\tau}\hat{a}^\dag_\xi\hat{a}_\lambda\hat{a}_\tau , \\
	i\dot{\hat{a}}_\nu&=-\sum_{\tau\in\mathcal{O}} R_{\nu\tau}\hat{a}_\tau,
\end{align}
\end{subequations}
so that 
\begin{subequations}
\begin{align}
\label{Helper1}
\langle\Psi|\hat{a}^\dag_\mu[\hat{H},\hat{a}_{\nu}]|\Psi\rangle = - \sum_{\tau\in\mathcal{O}} F_{\nu\tau}\rho_{\mu\tau},\\
\label{Helper2}
i\langle\Psi|\hat{a}^\dag_\mu\dot{\hat{a}}_\nu|\Psi\rangle = -\sum_{\tau\in\mathcal{O}} R_{\nu\tau}\rho_{\mu\tau}.
\end{align}
\end{subequations}
\section{Acceleration of Two-electron Term}\label{TwoElectronTerm}
The major computational bottleneck of Eq.~\eqref{defu} arises from evaluating the two-electron term $\bm{G}$. Instead of covering all possible non-zero multipoles, we truncate the spherical expansion in Eq.~\eqref{defG} at $\lambda_{\max}=\max(\lambda)$ and $\mu_{\max}=\max(|\mu|)$, as the contributions from higher-order terms are typically negligible. It significantly reduces the numerical cost.
 
To further improve efficiency without compromising accuracy, we use the methods described in Ref.~\cite{mccurdy2004implementation} to evaluate $G^{\mu\lambda}_{jk,nn'}$, which proves particularly efficient.
This approach extracts the radial component $G^{\mu\lambda}_{jk}(r)$ by solving the radial Poisson equation
\begin{equation}\label{RadialPoisson}
\bigg(\frac{d^2}{dr^2}-\frac{\lambda(\lambda+1)}{r^2}\bigg)rG(r) = -\frac{2\lambda+1}{r}\varrho(r),
\end{equation}
subject to proper boundary conditions. The complex-valued source term is given by
\begin{subequations}
\begin{align}
\varrho^{\mu\lambda}_{jk}(r) &= \sum_{mm',ll'}c^*_{jml}(r)C^{\mu\lambda}_{m'm,l'l}c_{km'l'}(r),\\
c_{iml}(r)	&=\sum_{n} c_{imln}B_n(r).
\end{align}
\end{subequations}
Only half of these terms need to be calculated due to the symmetry relation $G^{\mu\lambda}_{jk}(r)=[G^{-\mu\lambda}_{kj}(r)]^*$. After preparing $\bm{G}_{ij}$, the update vector $\bm{p}_{ijk} = \bm{G}_{ij}\bm{c}_k$ is calculated as follows 
\begin{subequations}
\begin{align}
\label{defp}
(\bm{p}_{ijk})_{\iota} &= \sum_{\mu\lambda n'}G^{\mu\lambda}_{ij,nn'} d^{\mu\lambda}_{kmln'},\\
\label{defd}
d^{\mu\lambda}_{kmln} &= \sum_{m'l'} C^{\mu\lambda}_{mm',ll'} c_{km'l'n}.
\end{align}
\end{subequations}
In a direct implementation, the numerical cost of the transforms scales with $O(N_{\rm a}\mu_{\max}\lambda_{\max}N_{\rm m}N_{\rm l}^2N_{\rm b})$ and $O(N_a^3\mu_{\max}\lambda_{\max}N_{\rm m} N_{\rm l} N_{\rm b} N_{\rm k})$ for Eqs.~\eqref{defd}\eqref{defp}, respectively, representing a severe computational bottleneck. 
This difficulty may be alleviated using the Discrete Variable Representation (DVR) for the angular variables $\eta=\cos\theta$ and $\varphi$ as discussed in Ref.~\cite{sukiasyan2001effect,haxton2007lebedev} and reviewed in Ref.~\cite{hochstuhl2014time}. Here, we choose Sukiasyan's method~\cite{sukiasyan2001effect}.
We start from the Gauss-Legendre quadrature of $C^{\mu\lambda}_{mm',ll'}$ on the interval $[-1,1]$, where
\begin{equation}\label{GLQ}
\begin{aligned}
C^{\mu\lambda}_{mm',ll'} &= \delta_{m',m-\mu}\sqrt{\frac{2}{2\lambda+1}} \times \\
&\sum_{g=1}^{N_{\rm G}} w_g N^m_l(\eta_g) N^\mu_\lambda(\eta_g)N_{l'}^{m'}(\eta_g),
\end{aligned}
\end{equation}
where $N^m_l$ is the normalized associated Legendre polynomials, $N_{\rm G}$ is the quadrature order, and $\eta_g$, $w_g$ are the quadrature points and weights, respectively. As Eq.~\eqref{GLQ} is numerically \textit{exact} when $2N_{\rm G}-1\ge2l_{\max}+\lambda_{\max}$, one may equivalently treat the two-electron terms in $\eta$'s DVR, where the orbitals and the multipole potentials are transformed by
\begin{subequations}
\begin{align}
\label{transformcFWD}
\tilde{c}_{imgn} &= \sum_{l}\chi^m_{gl}c_{imln}, \\
\label{transformcBWD}
c_{imln} 				&= \sum_{g}\chi^m_{gl}\tilde{c}_{imgn},\\
\label{transformG}
\tilde{G}^{\mu g}_{ij,nn'} &= \sum_{\lambda} \Lambda^{\mu}_{g\lambda}G^{\mu\lambda}_{ij,nn'}
\end{align}
\end{subequations}
with $\chi^m_{gl}=\sqrt{w_g}N^m_l(\eta_g)$, $\Lambda^\mu_{g\lambda}=\sqrt{\dfrac{2}{2\lambda+1}}N^\mu_\lambda(\eta_g)$. In $\eta$'s DVR, the calculation of $\tilde{\bm{p}}_{ijk}$ simplifies into
\begin{equation}
\label{CalcPijk}
(\tilde{\bm{p}}_{ijk})_{\tilde{\iota}} = \sum_{\mu m'} \delta_{m',m-\mu} \sum_{n'}\tilde{G}^{\mu g}_{ij,nn'}\tilde{c}_{km'gn'},
\end{equation}
where we have introduced the symbol $\tilde{\iota}=(m,g,n)$.

Similarly, adopting a uniform grid of $\varphi$ where $\varphi_f=\dfrac{2\pi f}{N_{\rm F}}$, the quadrature
\begin{equation}
\delta_{m',m-\mu}=\frac{1}{N_{\rm F}}\sum_{f=1}^{N_{\rm F}} e^{im\varphi_f}e^{-i\mu\varphi_f}e^{-im'\varphi_f}
\end{equation}
is \textit{exact} when $N_{\rm F}\ge|m'-m+\mu|$. We may simplify the two-fold summation over $\mu m'$ in Eq.~\eqref{CalcPijk} into one-fold by transforming the orbitals and the multipole potentials in $\varphi$'s DVR by
\begin{subequations}
\begin{align}
\label{transformcFWD2}
\mathring{c}_{ifgn} &= \frac{1}{\sqrt{N_{\rm F}}}\sum_{m}e^{-im\varphi_f}\tilde{c}_{imgn}, \\
\label{transformcBWD2}
\tilde{c}_{imgn} 				&= \frac{1}{\sqrt{N_{\rm F}}}\sum_{f}e^{im\varphi_f}\mathring{c}_{ifgn},\\
\label{transformG2}
\mathring{G}^{fg}_{ij,nn'} &= \sum_{\mu} e^{-i\mu\varphi_f}\tilde{G}^{\mu g}_{ij,nn'}.
\end{align}
\end{subequations}
With all the efforts above, the final expression for $\mathring{\bm{p}}_{ijk}$ is as simple as
\begin{equation}
\label{CalcPijk2}
(\mathring{\bm{p}}_{ijk})_{\mathring{\iota}} = \sum_{n'}\mathring{G}^{fg}_{ij,nn'}\mathring{c}_{kfgn'},
\end{equation}
where we have introduced the symbol $\mathring{\iota}=(f,g,n)$.

The procedure for calculating all $\bm{p}_{ijk}$ in the DVR framework now consists of three sequential steps. First, orbitals are transformed into the DVR basis using Eqs.~\eqref{transformcFWD} and~\eqref{transformcFWD2}, which involves independent groups of matrix-matrix products with complexity scaling as $O(N_{\rm a}N_{\rm m}N_{\rm l}N_{\rm G}N_{\rm b})$, alongside Fourier transforms scaling as $O(N_{\rm a}N_{\rm F}\log N_{\rm F}N_{\rm G}N_{\rm b})$. Second, multipole potentials are transformed into the DVR basis using Eqs.~\eqref{transformG}and ~\eqref{transformG2}, which has a complexity of $O(N_{\rm a}^2\mu_{\max}\lambda_{\max}N_{\rm G}N_{\rm b}N_{\rm k})$ for $\eta$ and $O(N_{\rm a}^2N_{\rm F}\log N_{\rm F}N_{\rm G}N_{\rm b}N_{\rm k})$ for $\varphi$, respectively. Third, the matrix-vector products defined in Eq.~\eqref{CalcPijk2} are performed, scaling as $O(N_{\rm a}^3 N_{\rm F}N_{\rm G} N_{\rm b} N_{\rm k})$. 
Using these procedures, the practical time consumption for evaluating $\bm{p}_{ijk}$ is comparable with the preparation of $\bm{G}_{jk}$. In molecular calculations with large $N_l$ and $N_m$, utilizing angular DVR typically yields several times faster execution than direct implementation methods, although the former necessitates increased computer memory for storing intermediate data and may require improved memory bandwidth. The cost can be further reduced by eliminating symmetry-induced redundancies.
For example, for $\rm C_s$ symmetry, half of the terms $G^{\mu g}_{ij,nn'}$ and $G^{fg}_{ij,nn'}$ vanish due to symmetry constraints on $\phi^*_i(\bm{r})\phi_j(\bm{r})$, allowing an additional halving of the computational cost by omitting these redundant calculations.

Subsequently, the vector $\bm{p}_{jkl}$ is used to compute $\bm{u}_i$ and the quantity $g_{ijkl}=\bm{c}^\dag_i\bm{p}_{jkl}$. 
The core-active interaction terms are computed similarly and benefit from further simplifications. 
For any frozen-core orbital $\phi_j$, all $\bm{G}_{jj}$ terms remain time-independent and need to be calculated only once.
In addition, frozen orbitals exhibit strong localization around the origin, effectively limiting summations over radial indices to a smaller subset.  
Other terms in Eq.~\eqref{defu}, such as the one-body integral $f_{ij}$, usually consume negligible time even if evaluated straightforwardly.

\end{document}